\documentclass[lettersize,journal]{IEEEtran}
\usepackage{amsmath,amsfonts}
\usepackage{algorithmic}
\usepackage{algorithm}
\usepackage{array}
\usepackage{ulem}
\usepackage[caption=false,font=normalsize,labelfont=sf,textfont=sf]{subfig}
\usepackage{textcomp}
\usepackage{stfloats}
\usepackage{url}
\usepackage{verbatim}
\usepackage{graphicx}
\usepackage{cite}

\usepackage{hyperref}
\hypersetup{
  colorlinks=true,
  linkcolor=blue,
  citecolor=blue,
  urlcolor=black
}
\usepackage{amssymb}
\usepackage{makecell}
\usepackage{picinpar}
\usepackage[caption=false,font=normalsize,labelfont=sf,textfont=sf]{subfig}
\usepackage{tabularx}
\usepackage{booktabs}
\usepackage[utf8]{inputenc}
\usepackage{colortbl}
\usepackage{soul}
\usepackage{multirow}
\usepackage{pifont}
\usepackage{color}
\usepackage{alltt}
\usepackage{enumerate}
\usepackage{siunitx}
\usepackage{epstopdf}
\usepackage{pbox}
\usepackage{titlesec}
\titlespacing\section{0pt}{12pt plus 4pt minus 2pt}{0pt plus 2pt minus 2pt}
\titlespacing\subsection{0pt}{12pt plus 4pt minus 2pt}{0pt plus 2pt minus 2pt}
\titlespacing\subsubsection{0pt}{12pt plus 4pt minus 2pt}{0pt plus 2pt minus 2pt}
\usepackage{geometry}
\begin{document}

\title{Dual-Envelope Constrained Nonlinear MPC for Distributed Drive Electric Vehicles Drifting Under Bounded Steering and Direct Yaw-Moment Control}

\author{Yurun Gan, Ziyu Song, Jing Yang, Zheng Lin, Jianuo Zhang, Tongtong Gu, Haitao Ding, Nan Xu, Por Lip Yee,  \\Wei Ni,~\IEEEmembership{Fellow,~IEEE}, and  Jun Luo,~\IEEEmembership{Fellow,~IEEE}
\thanks{This work was supported by the Joint Funds of the National Natural Science Foundation of China under Grant U1864206. (Corresponding author: Haitao Ding).}
\thanks{Yurun Gan, Ziyu Song, Jianuo Zhang, Tongtong Gu, Haitao Ding and Nan Xu are with State Key Laboratory of Automotive Simulation and Control, Jilin University, Changchun 130012, China.}
\thanks{Jing Yang and Por Lip Yee are with the Center of Research for Cyber Security and Network (CSNET), Faculty of Computer Science and Information Technology, Universiti Malaya, 50603 Kuala Lumpur, Malaysia.}
\thanks{Zheng Lin is with the Department of Electrical and Electronic Engineering, University of Hong Kong, Pok Fu Lam, Hong Kong, China.}
\thanks{Wei Ni is with the School of Engineering, Edith Cowan University, Perth, WA 6027, Australia.}
\thanks{Jun Luo is with the College of Computing
and Data Science, Nanyang Technological University, Singapore.}
% \thanks{Jing Yang and Por Lip Yee are with the Center of Research for Cyber Security and Network (CSNET), Faculty of Computer Science and Information Technology, Universiti Malaya, 50603 Kuala Lumpur, Malaysia (e-mail: s2147529@siswa-old.um.edu.my; porlip@um.edu.my).}
% \thanks{Z. Lin is with the Department of Electrical and Electronic Engineering, University of Hong Kong, Pok Fu Lam, Hong Kong, China (e-mail: linzheng@eee.hku.hk).}
% \thanks{W. Ni is with the School of Engineering, Edith Cowan University, Perth, WA 6027, Australia (email: wei/ni@ieee.org).}
% \thanks{J. Luo is with the College of Computing
% and Data Science, Nanyang Technological University, Singapore (e-mail: junluo@ntu.edu.sg).}

}

% The paper headers
\markboth{Journal of \LaTeX\ Class Files,~Vol.~14, No.~8, August~2021}%
{Shell \MakeLowercase{\textit{et al.}}: A Sample Article Using IEEEtran.cls for IEEE Journals}

\maketitle

\begin{abstract}
Distributed drive electric vehicles offer superior yaw moment control for autonomous drifting in extreme maneuvers. Conventional drift analysis constructs stability boundaries from open loop equilibria points and assumes a fixed envelope structure. However, coupling among control inputs reshapes the phase plane and shifts saddle point location, which can invalidate open loop envelopes when used for closed loop drifting. To address this issue, a saddle point coordinate model is established in this paper by combining a nonlinear tire model with the handling diagram and explicitly accounting for road adhesion coefficient, longitudinal velocity, front wheel steering angle, and additional yaw moment. Based on saddle point properties, an extended dual envelope framework is constructed in the phase plane of slip angle and yaw rate. Using the convergence tendency of state points toward saddle points under bounded control inputs, the outer envelope defines a recoverable set under constraints on front wheel steering angle and additional yaw moment. The inner envelope characterizes the non-drifting stability region associated with unsaturated tire forces. Finally, a nonlinear model predictive control (NMPC) controller is developed using the extended dual envelope constraint. Hardware-in-the-loop experiments show that, compared with NMPC without envelope constraints, the proposed method enables smoother convergence toward the drift saddle point, reduces the steady-state tracking errors of vehicle speed, sideslip angle, and yaw rate by 33.07\%, 71.18\%, and 31.27\%, respectively, and decreases the peak tracking error by 63.66\% under road-friction mismatch.

% compared with NMPC without envelope constraints, the proposed method enables smoother convergence toward the drift saddle point and improves robustness and tracking accuracy under road friction mismatch.
\end{abstract}

\begin{IEEEkeywords}
Distributed drive electric vehicles drifting, extended dual envelope, nonlinear model predictive control, saddle-point dynamics.
\end{IEEEkeywords}
\vspace{-5pt}
\section{Introduction}\label{section:1}

\IEEEPARstart{W}{ith} the rapid development of electrified and intelligent vehicle technologies~\cite{fang2024ic3m,lin2022channel,qu2025mobile,lin2024split,sun2025rrto}, distributed drive electric vehicles (DDEVs) have received increasing attention because of their fast torque response, flexible independent wheel actuation, and superior yaw moment control capability \cite{01,02,lin2022tracking}. These features make them well-suited for advanced motion control. Autonomous drifting is promising for extreme maneuvers, such as low adhesion driving, high speed obstacle avoidance, and aggressive path tracking near the handling limits \cite{02.5,03}. However, under sustained sideslip conditions, tire saturation and strong coupling among longitudinal, lateral, and yaw motions greatly complicate vehicle dynamics \cite{02.4,05}. If vehicle states cannot be maintained within the stable region, yaw divergence, motion instability, and even loss of control may occur \cite{06}. Therefore, it is of great theoretical and practical importance to develop a drift control framework for DDEVs that exploits actuation potential while preserving stability.

Vehicle phase diagram analysis and equilibrium point localization provide the foundation for controlled drifting. Drift control strategies can be tuned according to the vehicle state in the phase plane \cite{08}. Using a two degree-of-freedom (2DOF) model with nonlinear tire characteristics, Voser et al. \cite{09} showed that the drift equilibrium at large sideslip angles is a locally unstable saddle point. Subsequent studies extended this line of work from local equilibrium analysis to the global motion of unstable vehicles \cite{10}. Farroni et al. \cite{12} determined steady-state equilibria in a handling-diagram framework, while Hu et al. \cite{13} showed that unstable equilibria may still be surrounded by stable limit cycles, indicating that local stability alone cannot fully explain post-critical drifting behavior \cite{14}. 

More detailed four-wheel models further revealed the complexity of drifting equilibria. Milani et al. \cite{15} identified additional equilibrium pairs and classified them as primary and secondary drifting points, highlighting differences in drift mechanisms between rear-wheel-drive and four-wheel-drive vehicles. Tian et al. \cite{05} analyzed drift equilibria, all-wheel saturation, and deep sideslip using steady-state equations of a four-wheel model. However, most existing studies have primarily focused on locating the saddle point, with limited attention paid to the factors affecting its position and its variation under control inputs.

Phase plane analysis is widely used to evaluate vehicle lateral stability \cite{16}. It provides an intuitive description of trajectory convergence toward equilibrium points and helps construct stable envelopes for control design \cite{17}. Common representations include the slip angle-yaw rate ($\beta-r$) phase plane \cite{18}, energy phase plane \cite{41}, the front and rear tire slip angles phase plane \cite{20}, and the $\beta-\dot{\beta}$ phase plane \cite{21}, among which the $\beta-r$ phase plane is most widely used in drift control.

Based on drift equilibria, Bobier-Tiu et al. \cite{22} used phase-diagram isolines to define a stable envelope and shape the phase portrait so that the drifting system remained near the boundary, thereby reducing boundary violation and control chattering. However, conventional drift control mainly depends on steering and rear-wheel-drive actuation, which is limited by underactuation. To overcome this limitation, Lenzo et al. \cite{044} introduced front-wheel braking to enlarge the set of reachable drift equilibria, while Goh et al. \cite{24} proposed the maximum phase recovery envelope to identify uncontrolled regions in the phase plane. In addition, Yu et al. \cite{25} introduced a drift velocity lower bound for unified motion control in four-wheel-drive obstacle avoidance.

Nevertheless, most existing drift boundaries are established from open-loop phase-plane analysis and do not explicitly consider the effect of control inputs on the boundaries. Since coupled inputs such as steering angle and additional yaw moment may shift the saddle point, the corresponding attraction regions and boundaries should be reconsidered for drift control.

%The drift boundary settings described above are mainly derived from analysis of the saddle point location in the phase plane under open loop conditions, which means that the effect of control inputs on the saddle point location is not considered. During drifting, the influence of coupled control inputs, such as steering angle and additional yaw moment, on the saddle point location remains unclear. Therefore, the corresponding regions of attraction and drift boundaries in the vicinity of the saddle points must be delineated again to serve as constraints in controller design.

To fully exploit the vehicle dynamics phase diagram, the control method should incorporate phase-diagram-based constraints. Song et al. \cite{06} defined a safety boundary for steady-state drifting and imposed it as a hard constraint in iterative LQR for safe trajectory tracking. Model Predictive Control (MPC) is widely used in vehicle lateral stability control because it can explicitly handle system constraints while optimizing multiple objectives over a receding horizon. To address model mismatch under extreme conditions, Zhao et al. \cite{30} embedded sparse Gaussian process models into a two-layer MPC framework to improve control performance near drift equilibria. 

Since drifting usually occurs in the tire saturation region and involves strongly nonlinear vehicle dynamics \cite{31}, \cite{32}, Nonlinear Model Predictive Control (NMPC) is more suitable for drift control. Stano et al. \cite{33} incorporated drifting into a unified NMPC framework for emergency path tracking, while Zhao et al. \cite{34} introduced maximum phase recovery envelope (MPRE) constraints into NMPC to enlarge the recoverable maneuvering region. Weber et al. \cite{35} further developed an NMPC framework for online drift trajectory generation and tracking using a more realistic vehicle model. In addition, Chen et al. \cite{36} adopted MPC in DDEVs and demonstrated improved performance in extreme obstacle avoidance and path following.

However, the above methods still have two drawbacks in drifting. They often neglect boundary shifts caused by control inputs, and they usually focus only on the outer instability envelope while ignoring the inner stability region that may hinder convergence to the desired drift state.

\begin{figure*}
  \captionsetup{justification=raggedright, singlelinecheck=false}
  \centering
  \includegraphics[width=0.9\textwidth]{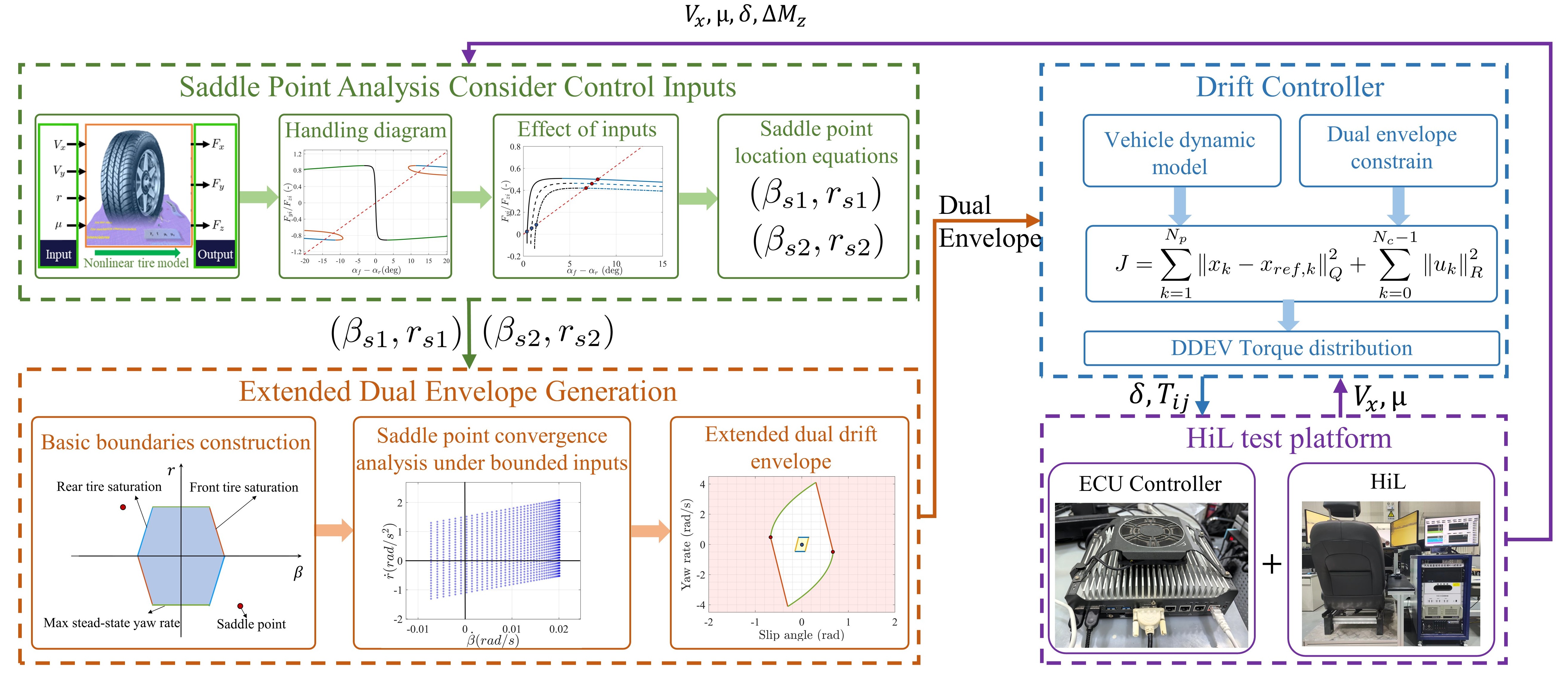}
  \caption{The proposed system architecture. This paper analyzes the factors affecting the saddle-point position and derives its governing equations under different input conditions. Based on the state derivatives near the saddle point, a control-input-aware dual drift envelope is constructed. An NMPC strategy is then developed for envelope regulation, and its effectiveness is validated on a hardware-in-the-loop (HiL) test platform.}\vspace{-20pt}
  \label{The proposed system architecture.}
\end{figure*}

Motivated by the above issues, this paper investigates drift dynamics of a DDEV using a single-track 3DOF vehicle model that incorporates coupled control inputs of front wheel steering angle and additional yaw moment. The architecture of the extended dual drift envelope is shown in Fig. \ref{The proposed system architecture.}. The main contributions of this paper are as follows.

\begin{enumerate}
  \item To address open the loop envelope failure during drifting caused by control input induced saddle point displacement, an input-coupled saddle-point coordinate equation is established for drifting by combining a nonlinear tire model with the handling diagram and incorporating the effects of steering angle and additional yaw moment. This extends conventional phase-plane analysis from fixed open-loop saddle points to variable saddle points under coupled control inputs.
  % a saddle point coordinate equation is established by combining a nonlinear tire model and explicitly accounting for control inputs. Saddle points are parameterized explicitly using fitted functions, which extends conventional phase diagram analysis based on fixed saddle points and open loop boundaries to an input coupled framework with variable saddle points.
  \item An extended dual-envelope constraint is proposed for drifting under bounded steering and yaw-moment control. A convergence index based on the phase-plane state-derivative direction is introduced to identify whether a state can be driven to the saddle point, thereby defining the outer recoverable boundary and the inner non-drifting stability boundary.
  
  % Considering coupled control inputs of front wheel steering angle and additional yaw moment, an extended dual envelope constraint is proposed. A convergence index based on changes in the direction of the phase plane state derivative is introduced to determine whether a state point can converge to a saddle point under bounded control inputs. Based on this index, the outer envelope boundary is constructed to distinguish controllable and uncontrollable regions. The inner envelope boundary is constructed from the non drifting stability region associated with unsaturated tire forces.
  \item A discrete NMPC controller incorporating the extended dual-envelope constraint is developed for DDEVs drift control. Experiments show that, compared with NMPC without envelope constraints, the proposed method improves tracking accuracy and robustness.
  
  % A discrete NMPC is developed, and the extended dual envelope constraint is incorporated into the optimization solver. The proposed constraints are evaluated using hardware-in-the-loop (HiL) experiments and are compared with controllers that do not use the dual envelope constraint. The results show that the proposed extended dual envelope NMPC controller improves tracking accuracy and robustness.
\end{enumerate}

%The detailed architecture of the extended dual drift envelope is illustrated in Fig. \ref{The proposed system architecture.}. This study investigates the factors influencing the saddle-point position in vehicle dynamics and derives the corresponding equations under different input conditions. Based on this, a drift double-envelope is constructed using state derivatives in its vicinity while explicitly incorporating control inputs. Furthermore, a nonlinear model predictive control strategy is developed to regulate the double envelope. The effectiveness and feasibility of the proposed method are finally validated through experiments conducted on a HiL test platform.

This paper is structured as follows. Section \ref{section:2} establishes the vehicle system model and the nonlinear tire model. Section \ref{section:3} introduces the analysis and fitting of the saddle point dynamics of the controlled input to obtain the saddle point position equations. Section \ref{section:4} presents the extended dual drift envelope design considering the control input. Section \ref{section:5} presents the design of the drift NMPC controller. Section \ref{section:6} carries out the experimental verification. Finally, Section \ref{section:7} provides the concluding remarks of this work.
\vspace{-10pt}
\section{System Model}\label{section:2}

%To capture the nonlinear dynamic characteristics of a drifting vehicle while maintaining computational tractability, this study adopts a single-track 3DOF vehicle model, together with a nonlinear tire model. The vehicle model includes an additional yaw moment, and the tire nonlinearity is represented using the UniTire-Ctrl model.

\vspace{-0pt}
\subsection{Vehicle Dynamic Model}
To represent vehicle states during drifting more accurately and apply appropriate torque for drift control, this paper adopts a 3DOF vehicle model that includes an additional yaw moment. The specific vehicle model is shown below:
\begin{equation}
    \begin{cases}
        \dot{V}_x = \dfrac{F_{xr} - F_{yf} \sin\delta}{m} + r V_y ;\\[8pt]
        \dot{\beta} = \dfrac{F_{yr} + F_{yf} \cos\delta}{m V_x} - r ;\\[8pt]
        \dot{r} = \dfrac{l_f F_{yf} \cos\delta - l_r F_{yr} + \Delta M_z}{I_z},
    \end{cases}
\end{equation}
where $V_x$ and $V_y$ denote the longitudinal and lateral vehicle speeds, respectively; $\beta$ denotes the sideslip angle; $r$ denotes the yaw rate; $\delta$ denotes the front wheel steering angle; $m$ denotes the vehicle mass; $I_z$ denotes the yaw moment of inertia; $\Delta M_z$ denotes the additional yaw moment about the $z$ axis; $l_f$ and $l_r$ denote the distances from the front and rear axles to the center of gravity, respectively; $F_{yf}$ denotes the lateral force at the front axle; $F_{xr}$ and $F_{yr}$ denote the longitudinal and lateral forces at the rear axle, respectively.
\vspace{-10pt}
\subsection{Path Tracking Model}
\begin{figure}
  \captionsetup{justification=raggedright, singlelinecheck=false}
  \centering
  \includegraphics[width=0.8\linewidth]{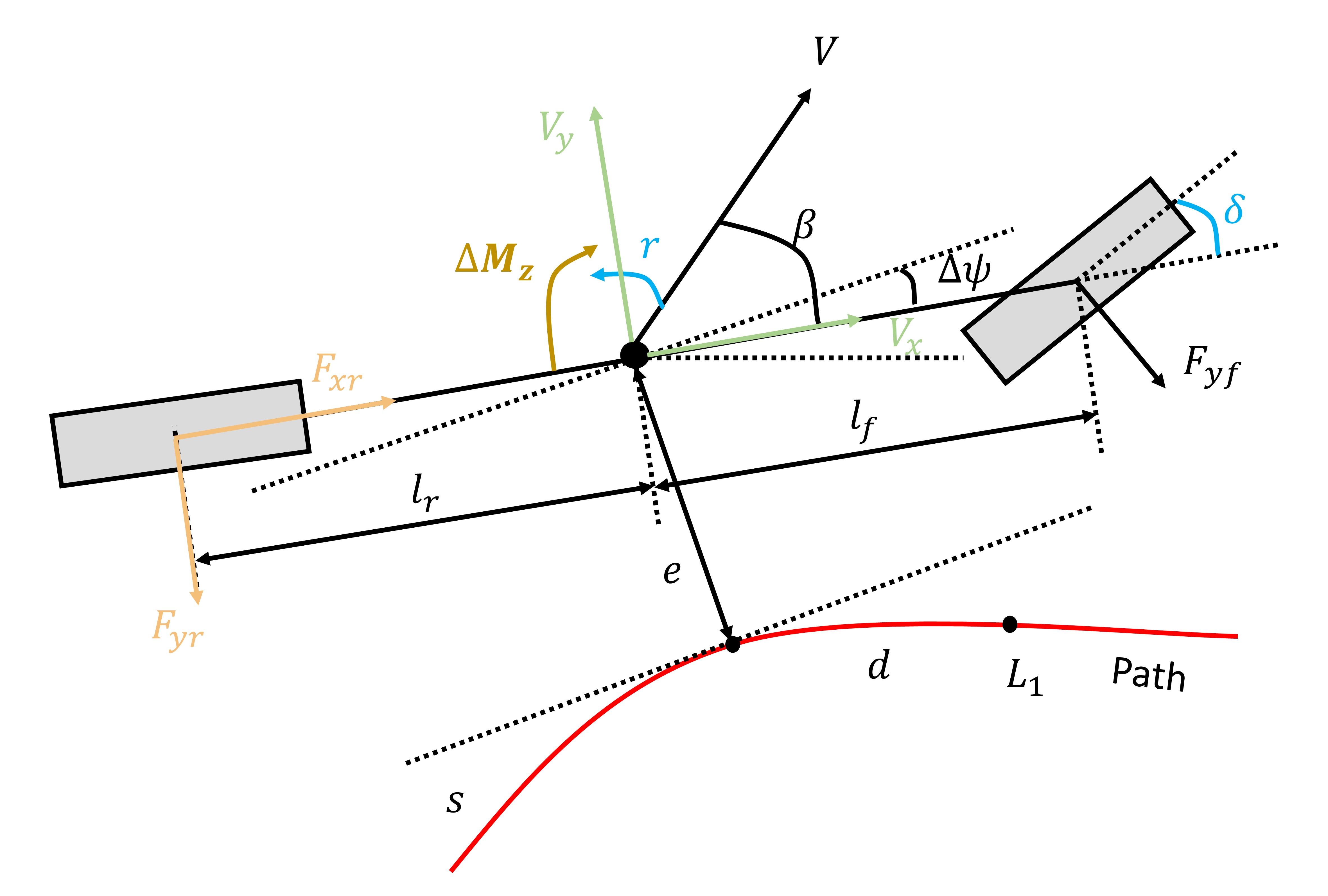}
  \caption{Single track model with reference path.}\vspace{-20pt}
  \label{Single track model with reference path.}
\end{figure}

Fig. \ref{Single track model with reference path.} presents a schematic of the reference dynamics model for a vehicle drifting along a prescribed path. The corresponding path tracking model is formulated in the Frenet coordinate system, and the governing equations are given below:
\begin{align}
	\begin{cases}
    \dot{e} = V \sin(\beta + \Delta \psi) 
                        = V_y \cos \Delta \psi + V_x \sin \Delta \psi; \\[6pt]
    \Delta \dot{\psi} = r - \kappa \dot{s}; \\[6pt]
    \dot{s}           = \frac{V \cos(\beta + \Delta \psi)}{1 - \kappa e} 
                        = \frac{V_x \cos \Delta \psi - V_y \sin \Delta \psi}{1 - \kappa e},
	\end{cases}
\end{align}
where $e$ denotes the distance between the vehicle center of gravity and the nearest point on the reference path, $s$ denotes the distance traveled along the current reference path, $\kappa$ denotes the current path curvature, $\Delta \psi$ denotes the yaw angle error relative to the reference yaw angle, and $V$ denotes the vehicle speed.
\vspace{-10pt}
\subsection{Nonlinear Tire Model}
Compared with the Magic Formula tire model, the UniTire model can better capture tire nonlinearities under combined slip conditions\cite{37}, \cite{38}. Therefore, this study adopts the UniTire-Ctrl tire model for drift control. In this model, the normalized resultant shear, denoted by $\bar F$, and the normalized longitudinal and lateral forces, denoted by $F_x$ and $F_y$, are expressed as follows:
\begin{equation}
\begin{cases}
\displaystyle
\bar F
= 1 - \exp\Bigl[-\phi - E\,\phi^2 - \bigl(E^2 + \tfrac{1}{12}\bigr)\phi^3\Bigr]; \\[0.5em]
\displaystyle
F_x
= \mu_x\,F_z\,\bar F\,
  \frac{\lambda_d\,\phi_x}
       {\sqrt{\bigl(\lambda_d\,\phi_x\bigr)^2 + \phi_y^2}}; \\[0.5em]
\displaystyle
F_y
= \mu_y\,F_z\,\bar F\,
  \frac{\phi_y}
       {\sqrt{\bigl(\lambda_d\,\phi_x\bigr)^2 + \phi_y^2}},
\end{cases}
\end{equation}
where $\phi_{x}$ and $\phi_{y}$ denote the normalized longitudinal and lateral ratios, respectively. The variable $\phi$ denotes the combined slip ratio. The parameter $E$ is a curvature coefficient that adjusts the shape of the resultant shear force curve in the transition region. The parameter $\lambda_{d}$ is a direction factor; $\mu_x$ and $\mu_y$ are the longitudinal and lateral friction coefficients, respectively; and $F_z$ denotes the vertical load acting on the tire.

The tire sideslip angles of the front and rear tires are expressed as follows:
\begin{equation}
\alpha_f
= \beta + {l_f \,r}/{V_x} - \delta; \quad \alpha_r
= \beta - {l_r \,r}/{V_x}.
\end{equation}
% \begin{equation}
% \begin{cases}
% \displaystyle
% \alpha_f
% = \beta + {l_f \,r}/{V_x} - \delta \\[0.8em]
% \displaystyle
% \alpha_r
% = \beta - {l_r \,r}/{V_x}
% \end{cases}
% \end{equation}

%To verify the accuracy of the UniTire-Ctrl tire model, this study conducted experiments on the same tires used in the test vehicle and compared the measured results with simulation outputs. The tires used in the experiments were Pirelli 205/45 R18. The comparison between experimental data and simulation results is shown in Fig. \ref{Validation of UniTire-Ctrl model with test vehicle}.
\begin{figure*}
      \captionsetup{justification=raggedright, singlelinecheck=false}
  \centering
  \includegraphics[width=0.85\linewidth]{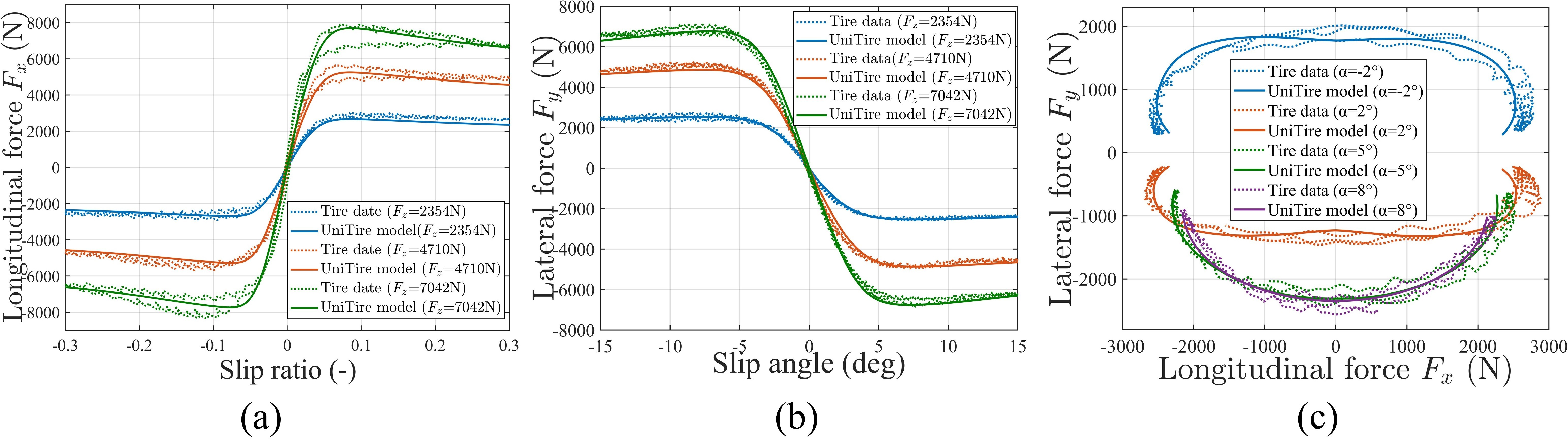}
  \caption{Validation of UniTire-Ctrl model. (a) The Pure longitudinal slip conditions. (b) The pure cornering conditions. (c) The combined slip conditions.}\vspace{-20pt}
  \label{Validation of UniTire-Ctrl model with test vehicle}
\end{figure*}

%Fig. \ref{Validation of UniTire-Ctrl model with test vehicle}a and Fig. \ref{Validation of UniTire-Ctrl model with test vehicle}b compare tire forces under pure longitudinal slip and pure sideslip conditions, respectively, for different vertical loads. Fig. \ref{Validation of UniTire-Ctrl model with test vehicle}c compares combined force responses at different sideslip angles under combined slip conditions. The results show that the fitted UniTire-Ctrl tire model agrees well with the test data, captures the relevant nonlinear characteristics, and is suitable for saddle point location fitting and envelope design under bounded input coupling.
To validate the UniTire-Ctrl tire model, Pirelli 205/45 R18 tires were tested under pure longitudinal slip, pure lateral slip, and combined slip conditions. As shown in Fig. \ref{Validation of UniTire-Ctrl model with test vehicle}, simulation results align closely with experimental data across varying vertical loads and slip angles. The model accurately captures nonlinear force responses, confirming its suitability for saddle point location fitting and envelope construction under bounded inputs.

\vspace{-10pt}
\section{Saddle Dynamics Analysis}\label{section:3}
%Considering tire nonlinearity during drifting, this section uses the Lyapunov indirect method to analyze vehicle stability under different tire operating states, examines the influence of control inputs, such as front wheel steering angle and additional yaw moment, on the saddle point location, and derives a saddle point location equation that accounts for relevant variables.

\subsection{Stability Analysis}
To analyze stability under extreme conditions while maintaining computational tractability, this study adopts a single-track 2DOF model with an additional yaw moment, and its dynamic equations are given as follows:
\begin{equation}
\begin{bmatrix}
\dot{\beta}\\
\dot{r}
\end{bmatrix}
= F(\beta,\, r,\, V_x,\, \delta,\, \mu, \Delta M_z)
=
\begin{bmatrix}
\dfrac{F_{yf}+F_{yr}}{mV_x}-r\\[6pt]
\dfrac{F_{yf}l_f - F_{yr}l_r+\Delta M_z}{I_z}
\end{bmatrix},
\end{equation}
where $\mu$ is the road adhesion coefficient.

\begin{table}[t]
\centering
\captionsetup{justification=centering}
\caption{Main Vehicle Parameters}
\label{tab:main_vehicle_parameters}
\footnotesize
\setlength{\tabcolsep}{3pt}
\renewcommand{\arraystretch}{0.95}
\begin{tabularx}{\columnwidth}{c >{\raggedright\arraybackslash}X c c}
\toprule
Symbol & Description & Value & Unit \\
\midrule
$m$   & Vehicle mass                        & 1720   & kg \\
$I_z$ & Yaw moment of inertia               & 1343.1 & kg$\cdot$m$^2$ \\
$l_f$ & Distance from CG to front axle      & 1.345  & m \\
$l_r$ & Distance from CG to rear axle       & 1.358  & m \\
$L$   & Wheelbase                           & 2.703  & m \\
$d$   & Track width                         & 1.660  & m \\
$R_e$ & Effective tire radius               & 0.32   & m \\
\bottomrule
\end{tabularx}\vspace{-20pt}
\end{table}

The parameters of the vehicle are shown in Table \ref{tab:main_vehicle_parameters}. Neglecting longitudinal load transfer during motion, the static vertical loads on the front and rear axles are given as:
\begin{equation}
F_{zf} = \frac{m g l_r}{l_f + l_r}; \quad F_{zr} = \frac{m g l_f}{l_f + l_r},
\end{equation}
% \begin{equation}
% \left\{
% \begin{aligned}
% F_{zf} &= \frac{m g l_r}{l_f + l_r}\\
% F_{zr} &= \frac{m g l_f}{l_f + l_r}
% \end{aligned}
% \right.
% \end{equation}
where $g$ is the acceleration due to gravity.

\begin{figure}
  \centering
  \includegraphics[width=0.6\linewidth]{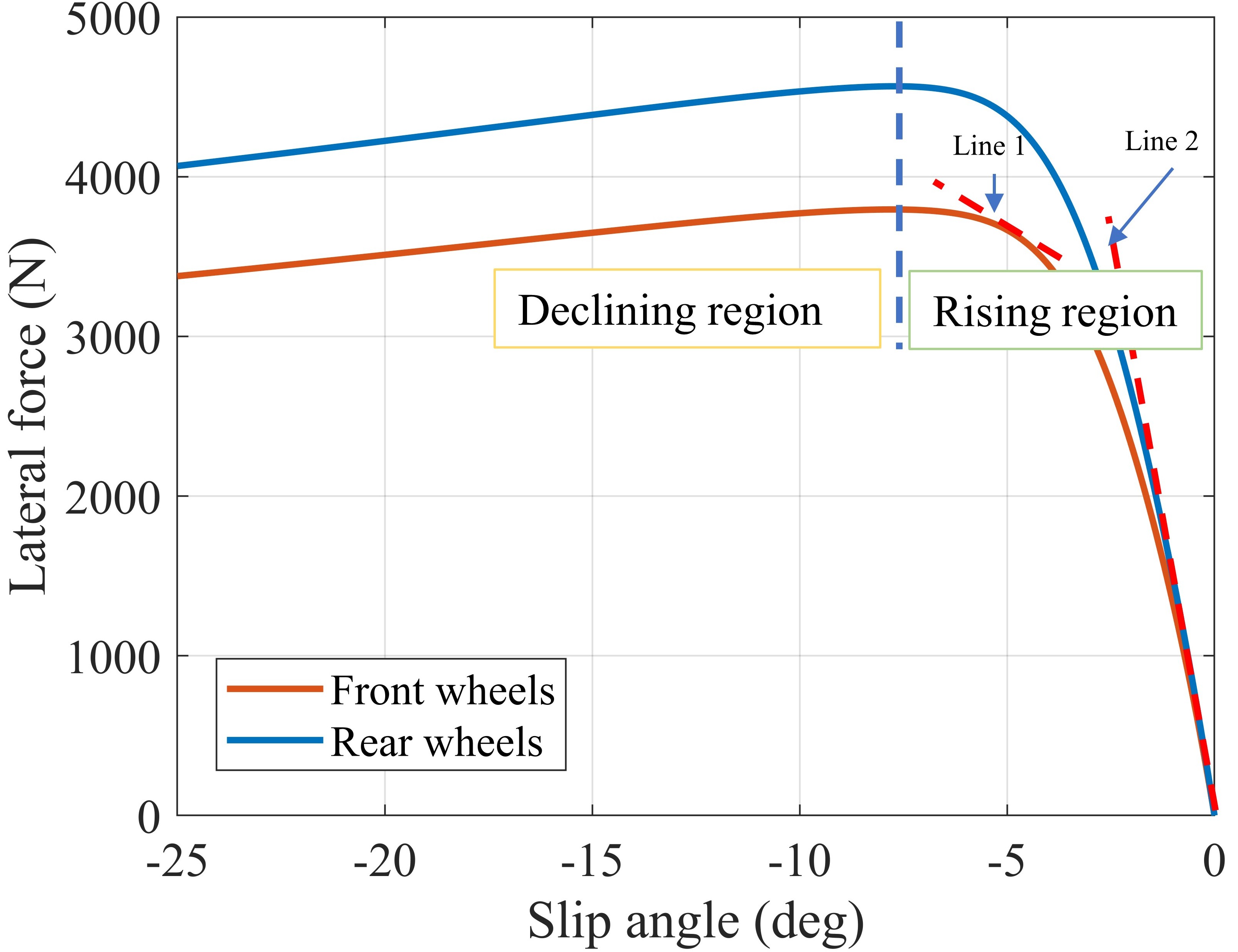}
  \caption{The lateral forces of the front and rear tires under combined slip conditions when $\mu=0.9$.}\vspace{-15pt}
  \label{The lateral forces of the front and rear tires under combined slip conditions when.}
\end{figure}

Through the UniTire-Ctrl tire model, the lateral force curves of the front and rear tires under combined slip conditions with $\mu = 0.9$ can be obtained, as shown in Fig. \ref{The lateral forces of the front and rear tires under combined slip conditions when.}.

Let $(\beta^*, r^*)$ denote an equilibrium point that satisfies $F(\beta, r, V_x, \delta, \mu, \Delta M_z) = 0$. Let Line 1 and Line 2 denote the tangent lines at this equilibrium point. Their slopes represent the tangential stiffness of the front and rear wheels at the equilibrium point, denoted by $C_{yf}^*$ and $C_{yr}^*$, respectively. To assess equilibrium stability, the Lyapunov indirect method is adopted. Taking partial derivatives with respect to the state variables yields the Jacobian matrix evaluated at the equilibrium point:
{\small
\begin{equation}
J(\beta^{*},r^{*})=
\begin{bmatrix}
\dfrac{C_{yf}^{*}+C_{yr}^{*}}{mV_x} &
\dfrac{C_{yf}^{*}l_f-C_{yr}^{*}l_r}{mV_x^{2}}-1 \\[2pt]
\dfrac{C_{yf}^{*}l_f-C_{yr}^{*}l_r}{I_z} &
\dfrac{C_{yf}^{*}l_f^{2}+C_{yr}^{*}l_r^{2}}{I_zV_x}
\end{bmatrix}.
\end{equation}
}

The equilibrium point is locally stable if $\mathrm{Tr}(J) < 0$ and $\mathrm{Det}(J) > 0$. Accordingly, system stability requires the following conditions.

\textbf{Case 1:} Both the front and rear tires operate in the rising region of the sideslip response, so that $C_{yf}^{*} < 0$ and $C_{yr}^{*} < 0$. If $C_{yf}^{*} l_f - C_{yr}^{*} l_r \ge 0$, the system is intrinsically stable and exhibits understeer or neutral steer characteristics. If $C_{yf}^{*} l_f - C_{yr}^{*} l_r < 0$, the system exhibits oversteer characteristics, and the stability condition is given by the following expression:
\begin{equation}
V_x < \sqrt{\frac{C_{yf}^{*}\,C_{yr}^{*}\,L^{2}}{m\left(C_{yr}^{*}l_r - C_{yf}^{*}l_f\right)}}.
\end{equation}

\textbf{Case 2:} The front tires operate in the declining region, whereas the rear tires operate in the rising region, so that $C_{yf}^{*} > 0$ and $C_{yr}^{*} < 0$. In the declining region, the slope is typically small, which implies $|C_{yr}^{*}| \gg |C_{yf}^{*}|$, and thus $\mathrm{Tr}(J) < 0$ holds. The remaining condition for system stability is given by the following expression:
\begin{equation}
V_x > \sqrt{\frac{C_{yf}^{*}\,C_{yr}^{*}\,L^{2}}{m\left(C_{yr}^{*}l_r - C_{yf}^{*}l_f\right)}}.
\end{equation}

\textbf{Case 3:} The front tires operate in the rising region, whereas the rear tires operate in the declining region, so that $C_{yf}^{*} < 0$ and $C_{yr}^{*} > 0$. In this case, $\mathrm{Det}(J) < 0$, which implies that the Jacobian has two eigenvalues of opposite sign. Therefore, the asymptotic stability condition is unsatisfied, and the equilibrium is a saddle point.

\textbf{Case 4:} The front and rear tires operate in the declining region, so that $C_{yf}^{*} > 0$ and $C_{yr}^{*} > 0$. In this case, $\mathrm{Tr}(J) > 0$ and $\mathrm{Det}(J) > 0$, which implies two positive eigenvalues and violation of the asymptotic stability condition.

The above analysis indicates that the saddle point locations are primarily determined by the lateral slip state of the tires. Saddle points occur when the front tires operate in the rising region and the rear tires operate in the declining region.
\vspace{-10pt}
\subsection{Handling Diagram Analysis}
From the above analysis, the saddle point occurs when the front tires operate in the rising region and the rear tires operate in the declining region. In addition, the handling diagram can be used to analyze the saddle point location of the vehicle \cite{12}. For saddle point, we can let $F(\beta, r, V_x, \delta, \mu, \Delta M_z) = 0$, the following relation is obtained:
\begin{equation}
\left\{
\begin{aligned}
\frac{a_y}{g} &= \frac{V_x^{2}}{g\,(l_f+l_r)}\left[\delta + (\alpha_f-\alpha_r)\right]; \\
\frac{a_y}{g} &= \frac{F_{yf}(\alpha_f)}{F_{zf}} = \frac{F_{yr}(\alpha_r)}{F_{zr}},
\end{aligned}
\right.
\end{equation}
where $F_{yf}(\alpha_f)$ and $F_{yr}(\alpha_r)$ denote the lateral forces that account for the additional yaw moment $\Delta M_z$. Specifically, $F_{yf}(\alpha_f) = F_{yf} + \Delta M_z / L$ and $F_{yr}(\alpha_r) = F_{yr} - \Delta M_z / L$.
  
\begin{figure}
  \centering
  \includegraphics[width=\linewidth]{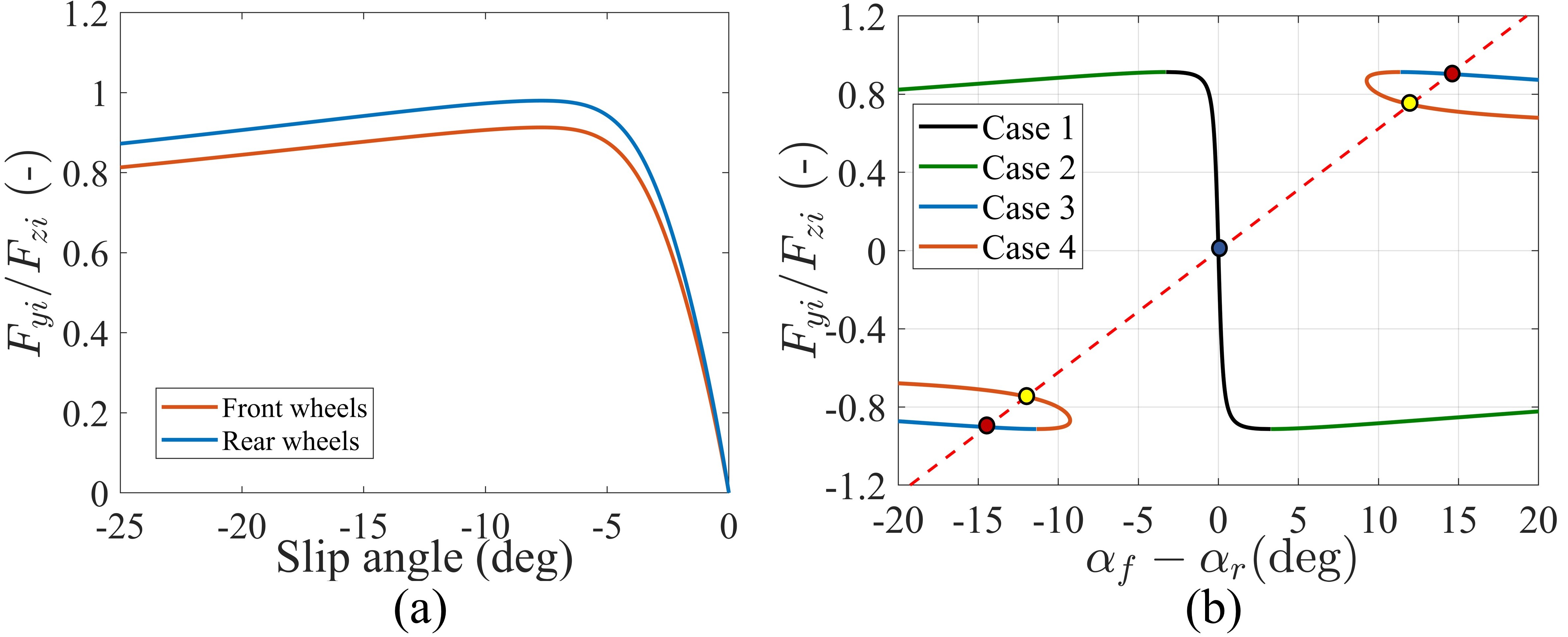}
  \caption{(a) The lateral sideslip characteristic curves of the front and rear wheels. (b) Handling diagram.}\vspace{-10pt}
  \label{(a) The lateral sideslip characteristic curves of the front and rear wheels. (b) Handling diagram.}
\end{figure}

As shown in Fig. \ref{(a) The lateral sideslip characteristic curves of the front and rear wheels. (b) Handling diagram.}, the handling diagram can be obtained for the operating condition $V_x = 60$ km/h, $\mu = 0.9$, and $\Delta M_z = 0$.
Fig. \ref{(a) The lateral sideslip characteristic curves of the front and rear wheels. (b) Handling diagram.}a shows the lateral sideslip characteristic curves of the front and rear tires; Fig. \ref{(a) The lateral sideslip characteristic curves of the front and rear wheels. (b) Handling diagram.}b shows the corresponding handling diagram constructed from these curves. The four cases correspond to the stability regimes associated with different front and rear tire operating regions discussed in the previous section. The red dashed line is defined by (10) and intersects the handling diagram. At the intersection between Case 1 and the dashed line, both the front and rear tires are unsaturated, which corresponds to the stable equilibrium point shown by the dark blue marker. At the red marker corresponding to the intersection between Case 3 and the dashed line, the front tire is unsaturated and the rear tire is saturated. Therefore, this intersection corresponds to the saddle point identified in the previous section, namely the drift equilibrium point. At the yellow marker corresponding to the intersection between Case 4 and the dashed line, both the front and rear tires are saturated, which implies a high risk of instability and loss of control. Therefore, this point is excluded from the considered operating range.

\begin{figure}
  \centering
  \includegraphics[width=0.55\linewidth]{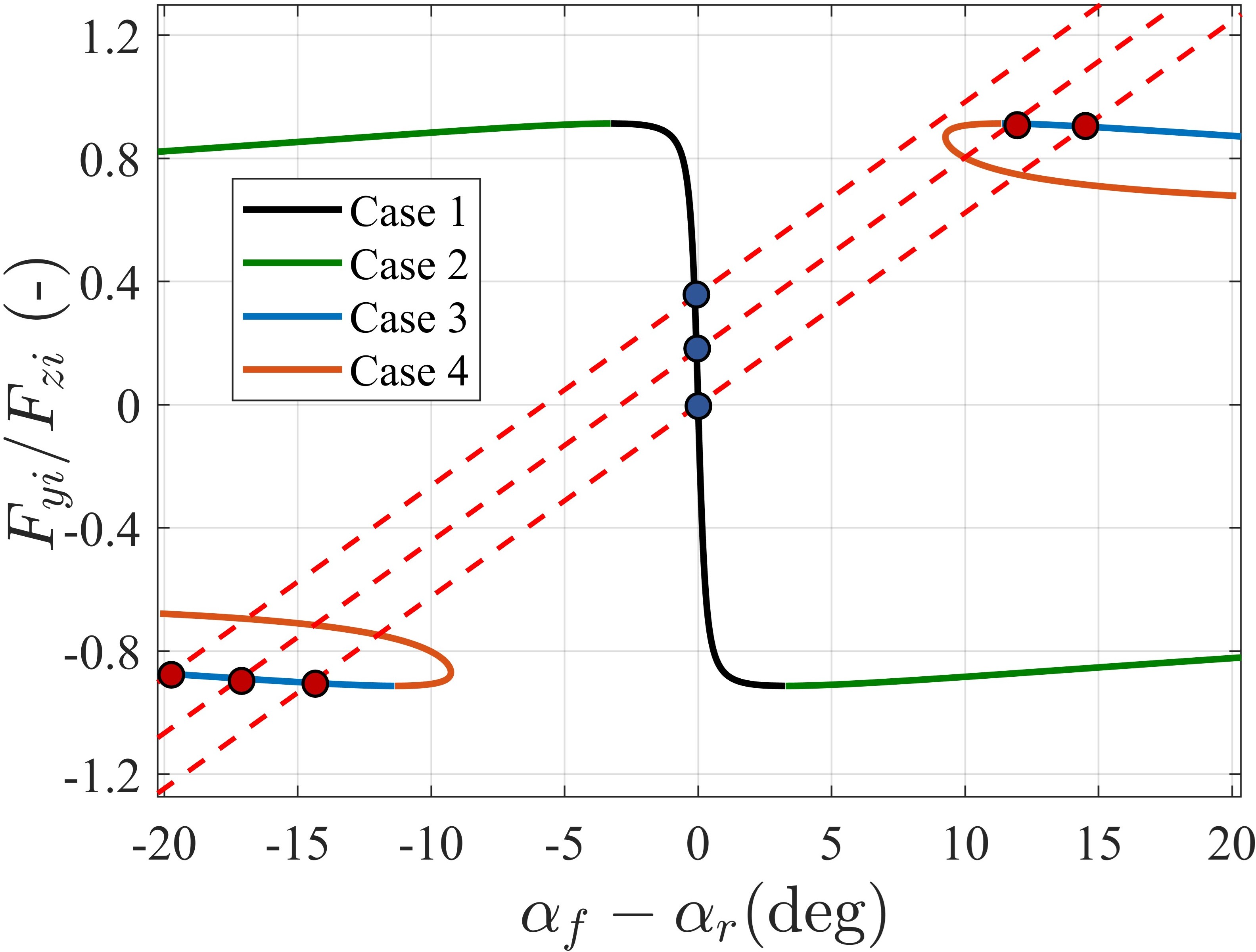}
  \caption{Saddle point positions at different front wheel steering angles.}\vspace{-10pt}
  \label{Saddle point positions at different front wheel steering angles.}
\end{figure}

As indicated by (10), changes in the front wheel steering angle cause corresponding changes in the vehicle saddle point location. The resulting saddle point variation is shown in Fig. \ref{Saddle point positions at different front wheel steering angles.}. As the front wheel steering angle increases, the saddle point shifts leftward, and the stable equilibrium point gradually approaches the right saddle point. When the front wheel steering angle increases beyond a critical value, the right saddle point disappears.

\begin{figure}[t]
  \centering
  \includegraphics[width=\linewidth]{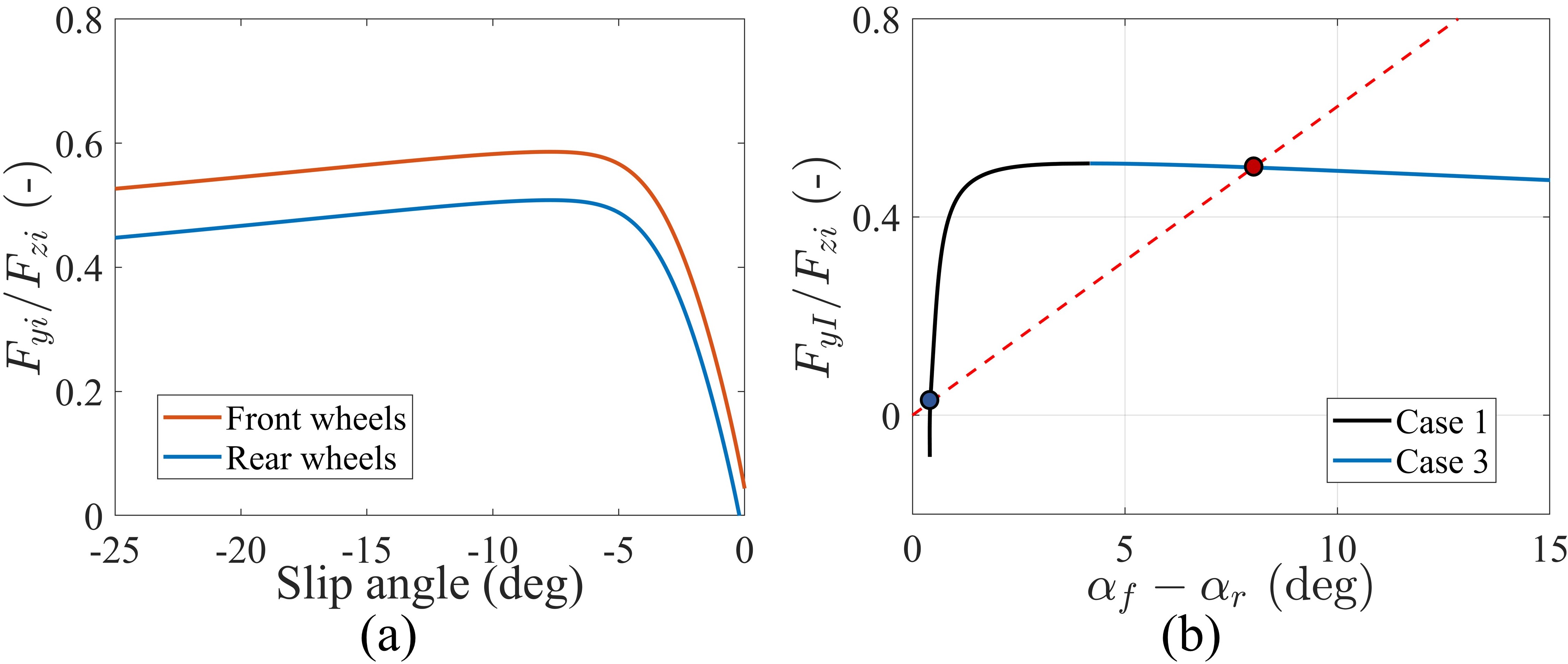}
  \caption{Handling diagram under $\Delta M_z=1000 Nm$. (a) The lateral sideslip characteristic curves. (b) Handling diagram.}\vspace{-10pt}
  \label{Handling diagram under.}
\end{figure}

\begin{figure}[t]
  \centering
  \includegraphics[width=0.55\linewidth]{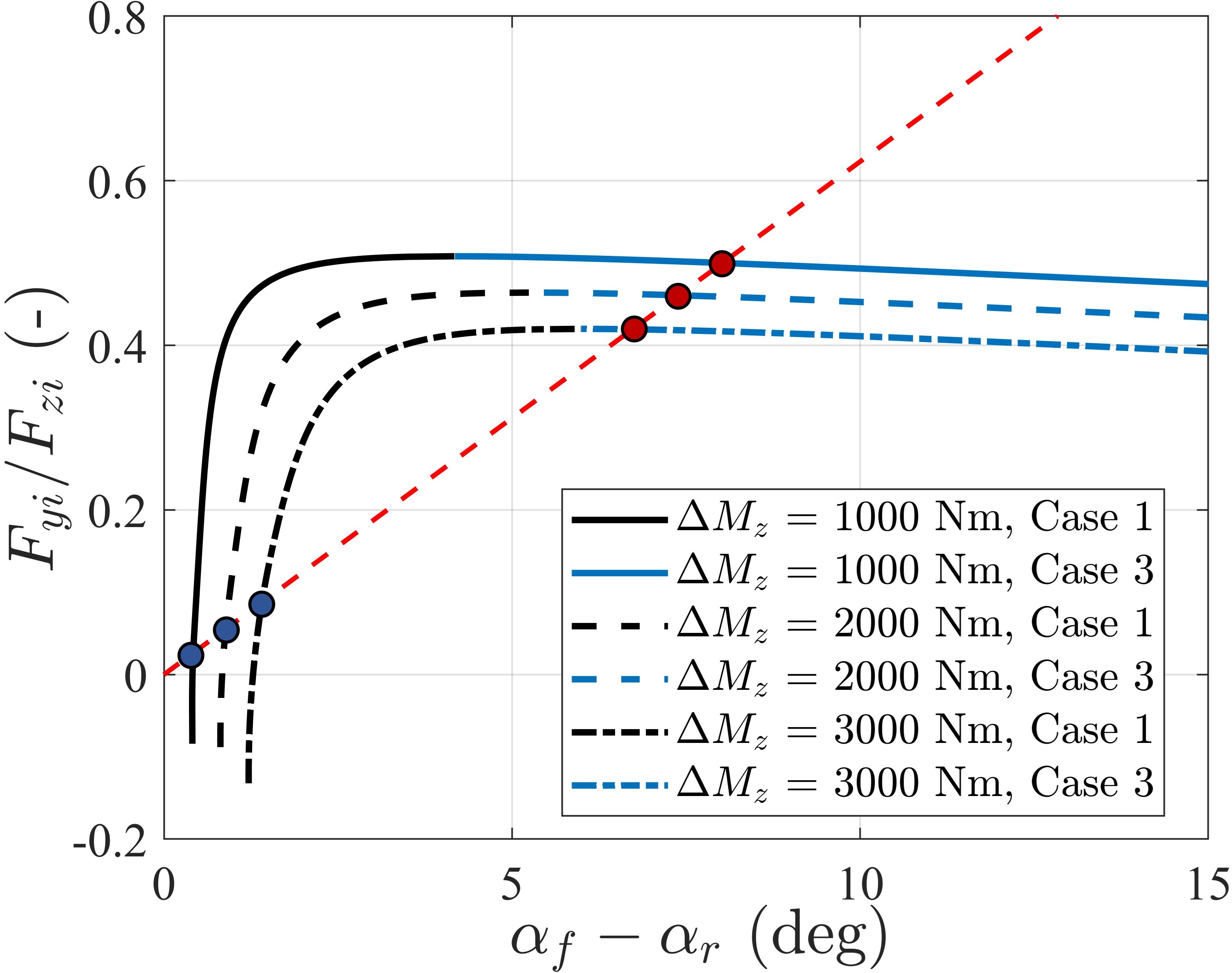}
  \caption{Saddle point position variation for different additional yaw moments.}\vspace{-15pt}
  \label{Saddle point position variation for different additional yaw moments.}
\end{figure}

In addition to the front wheel steering angle, additional yaw moment is included as a control input that influences the saddle point location. Fig. \ref{Handling diagram under.} shows the handling diagram for $V_x\!=\!60$ km/h, $\mu\!=\!0.5$, and $\Delta M_z\!=\!1000$ Nm. Under this additional yaw moment, the vehicle exhibits oversteer characteristics. In Fig. \ref{Handling diagram under.}b, the steady state solutions occur only in Case 1 and Case 3. Compared with Fig. \ref{(a) The lateral sideslip characteristic curves of the front and rear wheels. (b) Handling diagram.}, both the stable equilibrium point and the saddle point shift when an additional yaw moment is applied. Similarly, changes in the additional yaw moment shift the saddle point, as shown in Fig. \ref{Saddle point position variation for different additional yaw moments.}.

As shown in Fig. \ref{Saddle point position variation for different additional yaw moments.}, with vehicle speed, front wheel steering angle, and road adhesion coefficient held constant, increasing the additional yaw moment causes the stable equilibrium point to move toward the saddle point, while the saddle point shifts toward the lower left. The two points become progressively closer. This trend indicates that, beyond a critical additional yaw moment, the stable equilibrium point and the saddle point coalesce into a single equilibrium point.

\begin{figure}[t]
  \centering
  \includegraphics[width=0.6\linewidth]{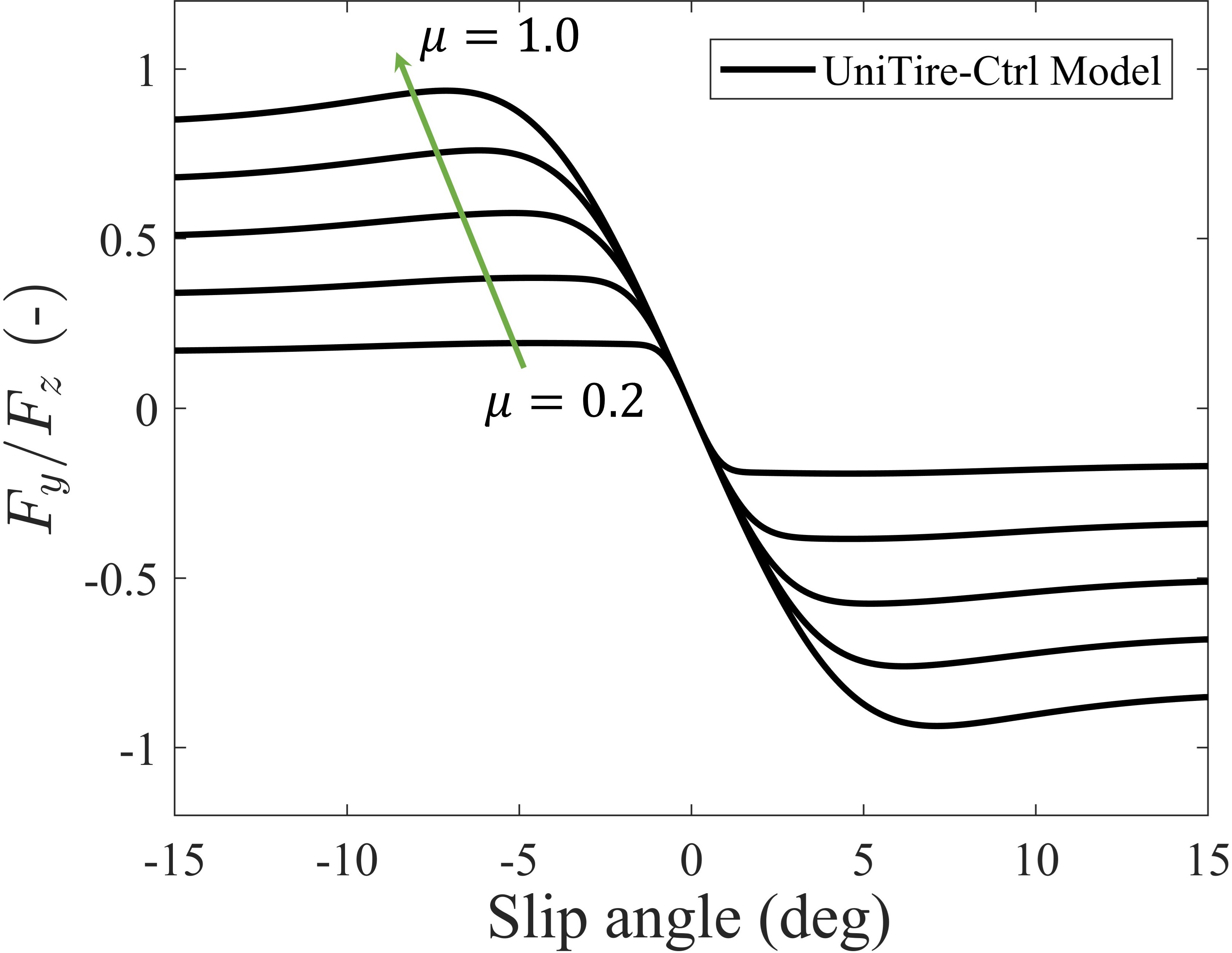}
  \caption{Normalized lateral force curve.}
  \label{Normalized lateral force curve.}\vspace{-10pt}
\end{figure}
\vspace{-10pt}
\subsection{Saddle Points Location Equation Derivation}
The above analysis of the front wheel steering angle and additional yaw moment indicates that variations in control inputs have a significant influence on saddle point location. In addition, parameters, such as vehicle speed and road adhesion coefficient, also affect saddle point location \cite{39}. This study derives a saddle point location equation that accounts for road adhesion coefficient, vehicle speed, front wheel steering angle, and additional yaw moment, as given by:
\begin{equation}
  \left\{
\begin{aligned}
r_{s1} &= \left(\frac{\mu g}{V_x} + f_1(\delta,\mu)\right)\cdot f_3(\Delta M_z,\mu,V_x) ;\\
r_{s2} &= \left(-\frac{\mu g}{V_x} + f_1(\delta,\mu)\right)\cdot f_3(\Delta M_z,\mu,V_x);
\end{aligned}
  \right.\vspace{-5pt}
\end{equation}
\begin{equation}
\small
\left\{
\begin{aligned}
\beta_{s1} &=
\Bigl(-\frac{l_f r_{s1}}{V_x}
+\bigl(-|\alpha_{f,\mathrm{sat}}|+\delta\bigr)\,
f_2(\delta,\mu,V_x)\Bigr)\,
f_4(\Delta M_z,\mu,V_x);\\
\beta_{s2} &=
\Bigl(-\frac{l_f r_{s2}}{V_x}
+\bigl(|\alpha_{f,\mathrm{sat}}|+\delta\bigr)\,
f_2(\delta,\mu,V_x)\Bigr)\,
f_4(\Delta M_z,\mu,V_x),
\end{aligned}
\right.
\end{equation}
where $\alpha_{f,sat}$ denotes the front tire slip angle at which the lateral force reaches its maximum, and it varies with the road adhesion coefficients, as shown in Fig. \ref{Normalized lateral force curve.}. The functional relationship between $\alpha_{f,sat}$ and $\mu$ is given by $\alpha_{f,sat} = p_1 \mu + p_2$. The variables $r_{s1}$ and $r_{s2}$ denote the yaw rates of the two saddle points, and $\beta_{s1}$ and $\beta_{s2}$ denote the corresponding slip angles. 
It is worth emphasizing that these saddle equilibria are not fixed features of the open-loop dynamics; rather, they should be treated as input-coupled and parameter-dependent equilibria, since admissible control actions reshape the underlying vector field and continuously shift the equilibrium locations. 
The functions $f_1$ and $f_2$ are saddle point location adjustment functions that fine tune the saddle point location according to changes in state and control inputs. The functions $f_3$ and $f_4$ characterize the influence of additional yaw moment on saddle point location, and are defined as follows:
\begin{equation}
\left\{
\begin{aligned}
f_1 &= p_3\frac{\delta}{\mu} \\
f_2 &= p_4 + p_5\,\delta\mu + p_6\,V_x \\
f_3 &= 1-\left(\frac{\Delta M_z}{\mu p_7\left(1-(p_8V_x+p_9)\right)}\right)^2 \\
f_4 &= 1-\left(\frac{\Delta M_z}{\mu p_{10}\left(1-(p_11V_x+p_{12})\right)}\right)^2
\end{aligned}
\right.
\end{equation}
where $p_1$ to $p_{12}$ are model parameters identified using the MATLAB cftool. Parameter identification is conducted by varying the control inputs and vehicle states, locating the corresponding saddle points, and fitting the data.
\vspace{-10pt}
\section{Extended dual drift stable envelope design considering the control inputs}\label{section:4}
%This section analyzes the drifting saddle point and its surrounding states using the traditional phase plane stability boundary, and evaluates the corresponding state derivative distributions under bounded control inputs. A convergence index derived from the state derivative direction is then introduced to assess whether a given phase plane region can be driven to converge to the saddle point under control action. Based on this criterion, a drift oriented, extended dual envelope is constructed, and its variation with vehicle speed and road adhesion coefficient is further investigated.\vspace{-10pt}

\subsection{Nonlinear Phase Plane Analysis}
\begin{figure}
  \centering
  \includegraphics[width=0.55\linewidth]{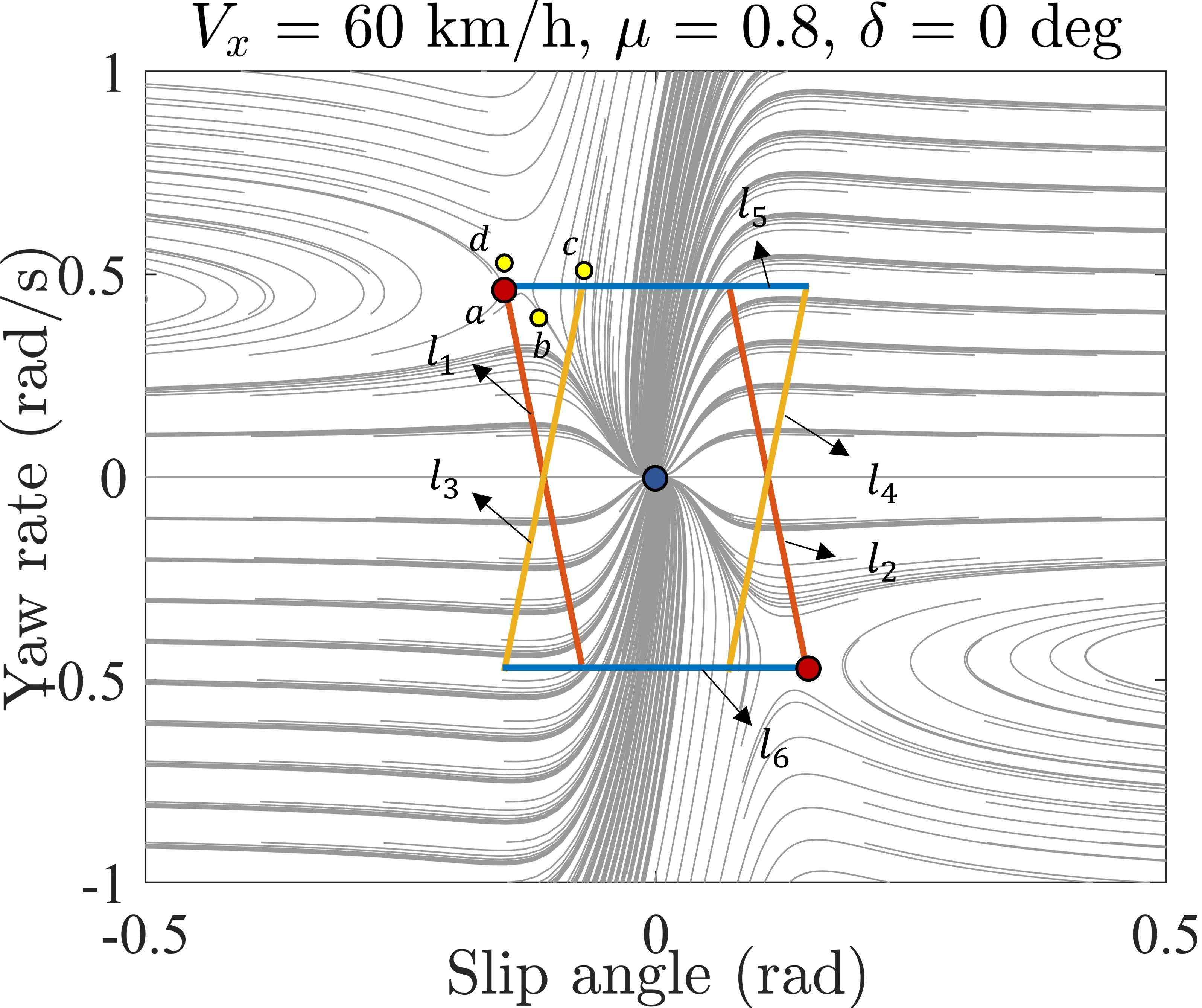}
  \caption{The stability boundaries without considering control inputs.}\vspace{-18pt}
  \label{The stability boundaries without considering control inputs.}
\end{figure}

For traditional vehicle dynamic stability boundaries, a typical boundary configuration is illustrated in Fig. \ref{The stability boundaries without considering control inputs.}. The boundary does not account for control inputs and represents an open loop boundary. Here, $l_1$ and $l_2$ denote the front tire saturation boundaries, $l_3$ and $l_4$ denote the rear tire saturation boundaries, and $l_5$ and $l_6$ denote the maximum steady state yaw rate boundaries achievable by the vehicle. However, these bounds do not incorporate control input characteristics and can therefore lead to inaccurate constraint enforcement in vehicle safety control.

\begin{figure}
  \centering
  \includegraphics[width=\linewidth]{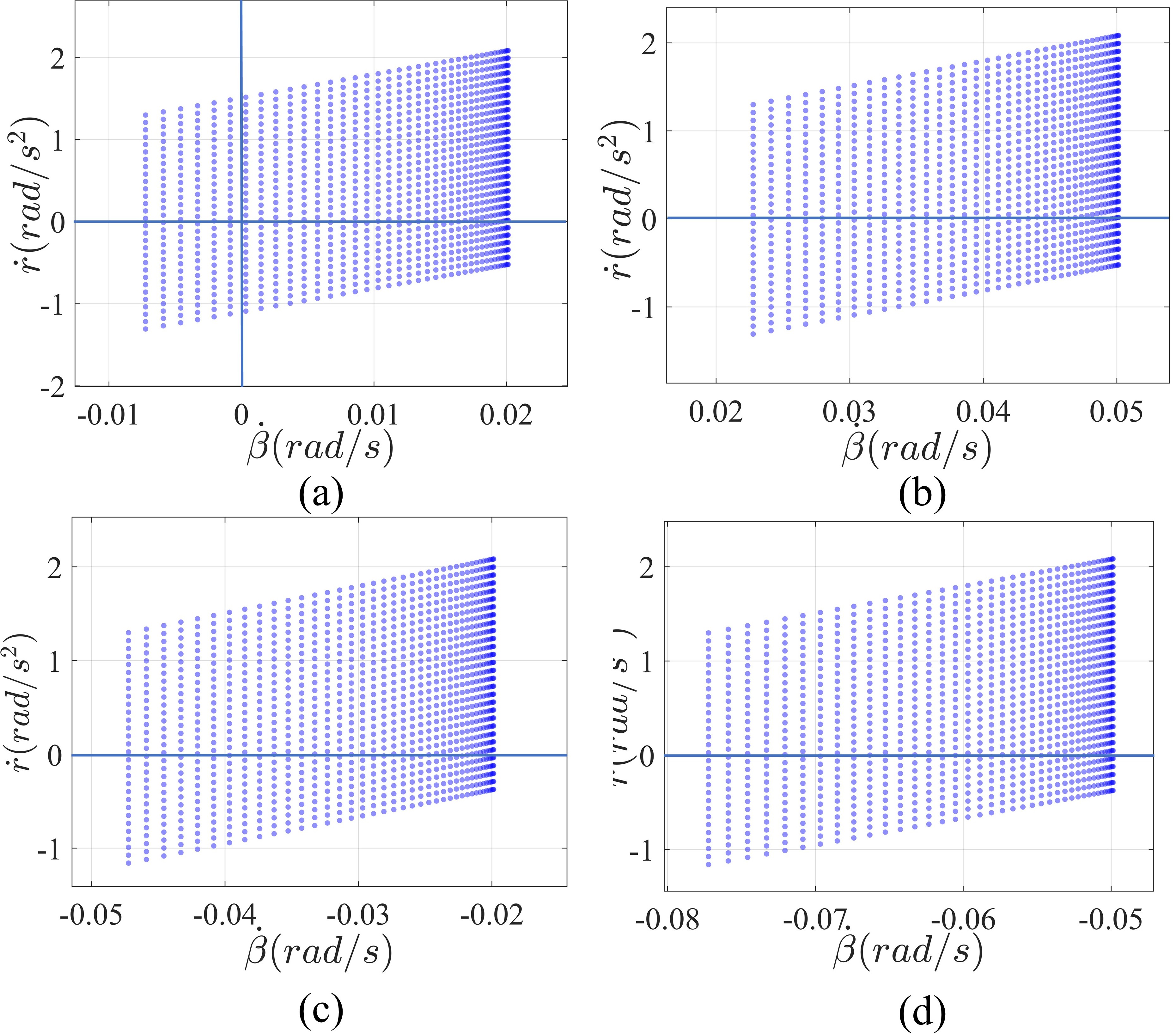}
  \caption{The state derivatives of saddle point and its surrounding points. (a) saddle point, $\beta=-0.14$ rad, $r=0.43$ rad/s. (b) $\beta=-0.10$ rad, $r=0.40$ rad/s. (c) $\beta=-0.05$ rad, $r=0.47$ rad/s. (d) $\beta=-0.145$ rad, $r=0.50$ rad/s.}\vspace{-15pt}
  \label{The state derivatives of saddle point and its surrounding points}
\end{figure}

In practice, the stability bounds described above are insufficient for drifting, particularly when the drift equilibrium lies outside the conventional open loop stability boundary. Therefore, designing a stability boundary suitable for drifting requires consideration of state derivative variations in the vicinity of the saddle point. As shown in Fig. \ref{The stability boundaries without considering control inputs.}, points b, c, and d are state points near the saddle point. Along the open loop phase trajectories, all three points move toward the saddle point but do not reach it. When the vehicle is subject to coupled front wheel steering angle and additional yaw moment, the phase plane state derivatives change, which alters the convergence behavior near the saddle point. Using bounded control inputs of $-0.5 \le \delta \le 0.5$ rad and $-3500$ Nm $\le \Delta M_z \le 3500$ Nm, the state derivatives at the saddle point (point a), and the surrounding points are computed separately, and the results are shown in Fig. \ref{The state derivatives of saddle point and its surrounding points}.

Fig. \ref{The state derivatives of saddle point and its surrounding points}a shows the state derivatives at the saddle point and at nearby state points when control inputs are applied.
The nullclines $\dot{\beta} = 0$ and $\dot{r} = 0$ intersect within the state derivative field, which indicates that coupled front wheel steering angle and additional yaw moment can stabilize the vehicle at the saddle point.
In Fig. \ref{The state derivatives of saddle point and its surrounding points}b, the condition $\dot{\beta}\beta < 0$ indicates that the control inputs drive $\beta$ toward zero and do not induce instability. In Fig. \ref{The state derivatives of saddle point and its surrounding points}c and Fig. \ref{The state derivatives of saddle point and its surrounding points}d, the condition $\dot{\beta}\beta > 0$ indicates that $\beta$ increases, although the growth rate differs between the two cases. Under suitable coupling of control inputs, a subset of the region containing points c and d can be driven toward point a and remain there. The remaining question is how to identify such regions.
\vspace{-10pt}
\subsection{Extended Dual Drift Stable Envelope Design}

During drifting, the vehicle operates in Case 3, in which the front tires remain unsaturated and the rear tires are saturated. Therefore, the front tire saturation boundary must be determined first to ensure that the front tires remain in the unsaturated regime during drifting. Taking the left saddle point as an example, the front tire boundary is given by:
\begin{equation}
\beta \geqslant  -\frac{l_f}{V_x} r+\delta_{min}+\alpha_{f,sat}.
\end{equation}

Similarly, the rear tires should be saturated during drifting. Therefore, the rear tire saturation boundary serves as an inner boundary; i.e., the vehicle must cross this boundary to enter the drift region. The rear tire saturation boundary is given by :
\begin{equation}
\beta \leqslant  \frac{l_r}{V_x} r+\alpha_{r,sat},
\end{equation}
where $\alpha_{r,sat}$ denotes the slip angle, at which the rear tire lateral force reaches its maximum value.

Considering the influence of control inputs, the maximum steady-state yaw rate is given by (11), and the specific equation is given as follows:
\begin{equation}
r \geqslant  \left(\frac{\mu g}{V_x} + f_1(\delta,\mu)\right)\cdot f_3(\Delta M_z,\mu,V_x).
\end{equation}

The safety bounds defined above describe only the limits of the stable region. A portion of the phase plane near the left saddle point in Fig. \ref{The stability boundaries without considering control inputs.} remains undetermined, namely the region containing points c and d. Although this region lies beyond the maximum steady state yaw rate boundary, states within it may still reach the saddle point because control inputs can modify the local convergence behavior. Therefore, this region is examined in greater detail below.

\begin{figure}
  \centering
  \includegraphics[width=\linewidth]{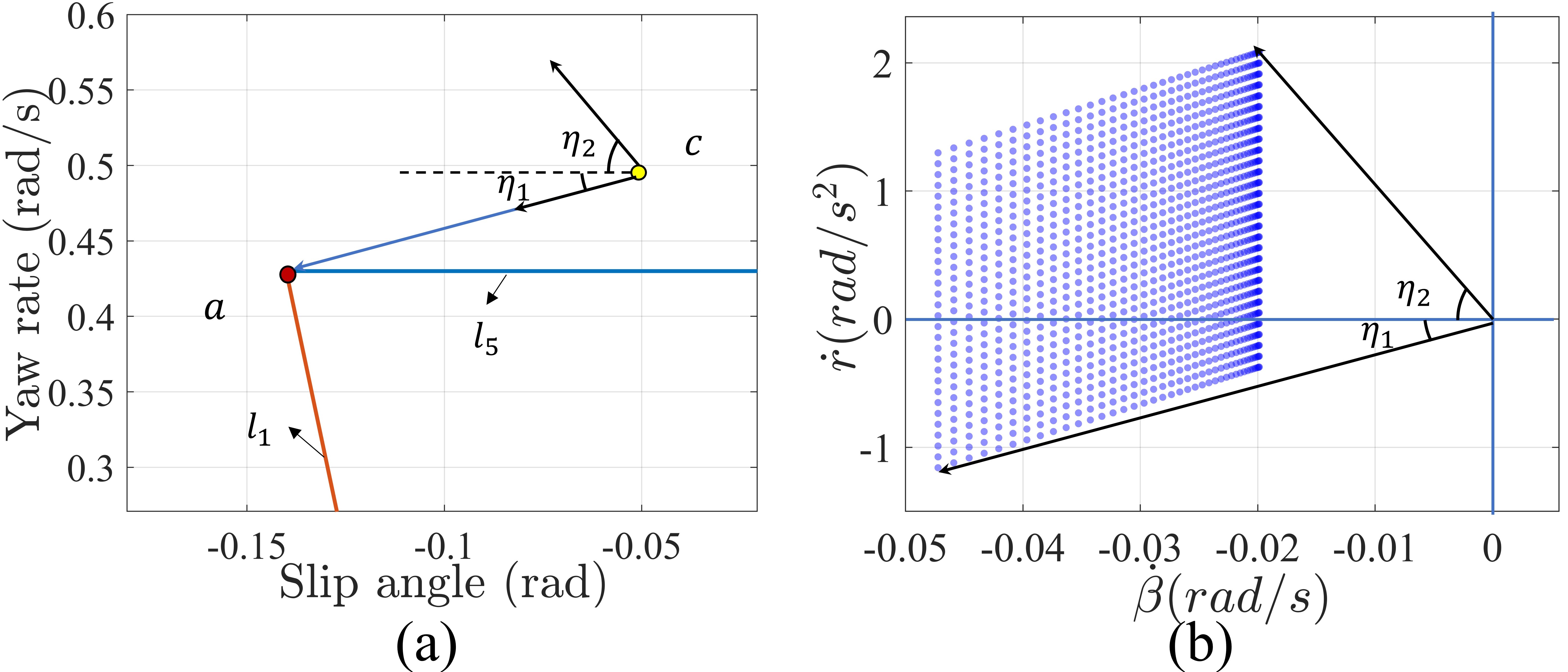}
  \caption{The degree of convergence from point c to the saddle point a under control inputs. (a) $\beta-r$ State direction (b) $\dot{\beta}-\dot{r}$ State derivative}\vspace{-20pt}
  \label{The degree of convergence from point c to the saddle point a under control inputs.}
\end{figure}

Inspired by \cite{38}, this section introduces an index that quantifies the ability of a vehicle state to converge to a saddle point under control inputs. As shown in Fig. \ref{The degree of convergence from point c to the saddle point a under control inputs.}a, $\eta$ is defined as the angle between the line connecting a state point to the saddle point in the $\beta$ to $r$ phase plane and the $\beta$ axis. The index $\eta$ characterizes the range, over which the direction of the state derivative can be altered by control inputs. Point c lies outside the maximum steady state yaw rate boundary. Under control inputs, its state derivative set is shown in Fig. \ref{The degree of convergence from point c to the saddle point a under control inputs.}b. The corresponding angular range is bounded by $\eta_1$ and $\eta_2$. When $\eta = \eta_2$, point c can be driven to point a by the control inputs. This result indicates that, even if states near the saddle point exceed the previously defined boundary, control inputs can still steer them to the saddle point and maintain stability, thereby enabling stable drifting.

Let $\eta_a$ denote the angle of the line connecting a state point near the saddle point to the saddle point $a$. From the above discussion, the convergence of a state point near the saddle point is governed by the maximum admissible control inputs. Therefore, whether a vehicle state near the saddle point can be driven to converge to the saddle point under control action can be determined by:
\begin{equation}
\eta_{\max} \geqslant \eta_a,
\end{equation}
where $\eta_a = \arctan\left(\frac{r - r_a}{\beta - \beta_a}\right)$, and $\eta_{max}$ denotes the maximum of $\eta$ achievable under bounded control inputs. When $\dot{r}$ attains its minimum, $\eta$ reaches its maximum, so $\eta_{max}$ can be expressed as follows:
\begin{equation}
\begin{aligned}
\eta_{\max } & =\arctan \left(\frac{\dot{r}_{\min }}{\dot{\beta}}\right) \\
& =\arctan \left(\frac{F_{y f \min } l_f-F_{y r \max } l_r+\Delta M_{z \min }}{I_z}\right. \\
& \left.\cdot \frac{m V_x}{F_{y f \min }+F_{y r \max }-r m V_x}\right)
\end{aligned}
\end{equation}
where $F_{yfmin}$ and $F_{yrmax}$ denote the front tire lateral force and rear tire lateral force, respectively, corresponding to the minimum of $\dot{r}$ under bounded control inputs in Fig. \ref{The degree of convergence from point c to the saddle point a under control inputs.}b. The parameter $\Delta M_{z\min}$ denotes the additional yaw moment corresponding to the minimum of $\dot{r}$ under bounded control inputs.

\begin{figure}
  \centering
  \includegraphics[width=0.75\linewidth]{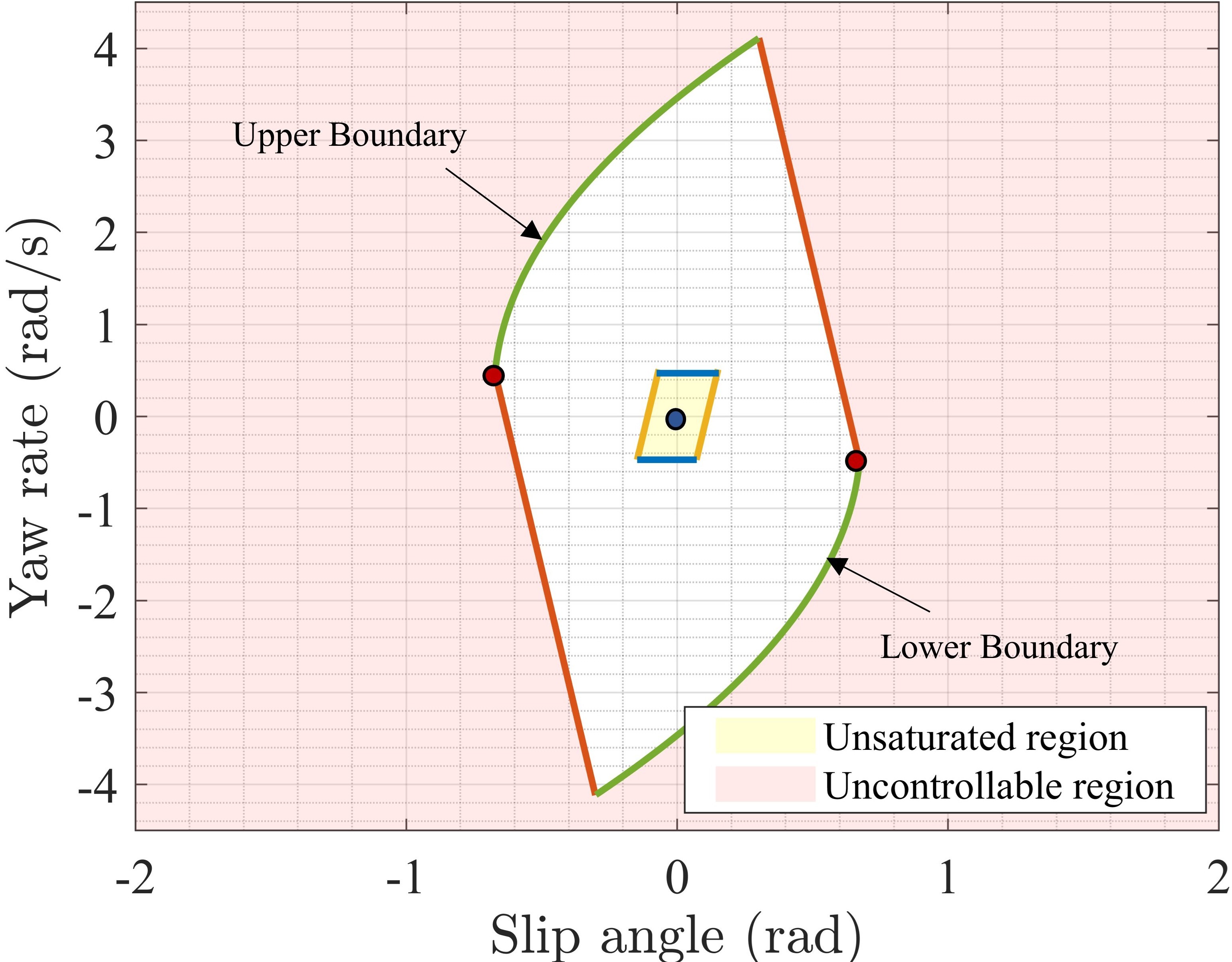}
  \caption{Considering the drift extended dual envelope under the influence of control inputs the vehicle speed is 60km/h and the road adhesion coefficient is $\mu=0.8$.}\vspace{-10pt}
  \label{Considering the drift double extended envelope under the influence of control inputs.}
\end{figure}

Based on the above calculations, an upper stable extended boundary near the saddle point that accounts for control inputs can be obtained, as shown in Fig. \ref{Considering the drift double extended envelope under the influence of control inputs.}. The left saddle point corresponds to the saddle point location under extreme control input conditions and does not represent all saddle points at $V_x = 60$ km/h. For the right saddle point, the envelope is constructed using the same method, and the resulting envelope segment forms the lower boundary.

The above curves form two closed envelopes. The inner envelope consists of the blue and yellow curves. The enclosed yellow region corresponds to a stable operating domain, in which neither tire saturation nor a yaw rate boundary is reached, and it therefore represents an unsaturated region. The outer envelope consists of the red and green curves. The red curve denotes the front tire saturation boundary, and the green curve denotes the boundary of states that can converge to the saddle point under bounded control inputs. The pink region outside the outer envelope corresponds to an uncontrollable domain. Therefore, the effective drift driving domain during drifting is the white region. At this stage, the extended dual envelope for drifting is established.

\begin{figure}
  \centering
  \includegraphics[width=1\linewidth]{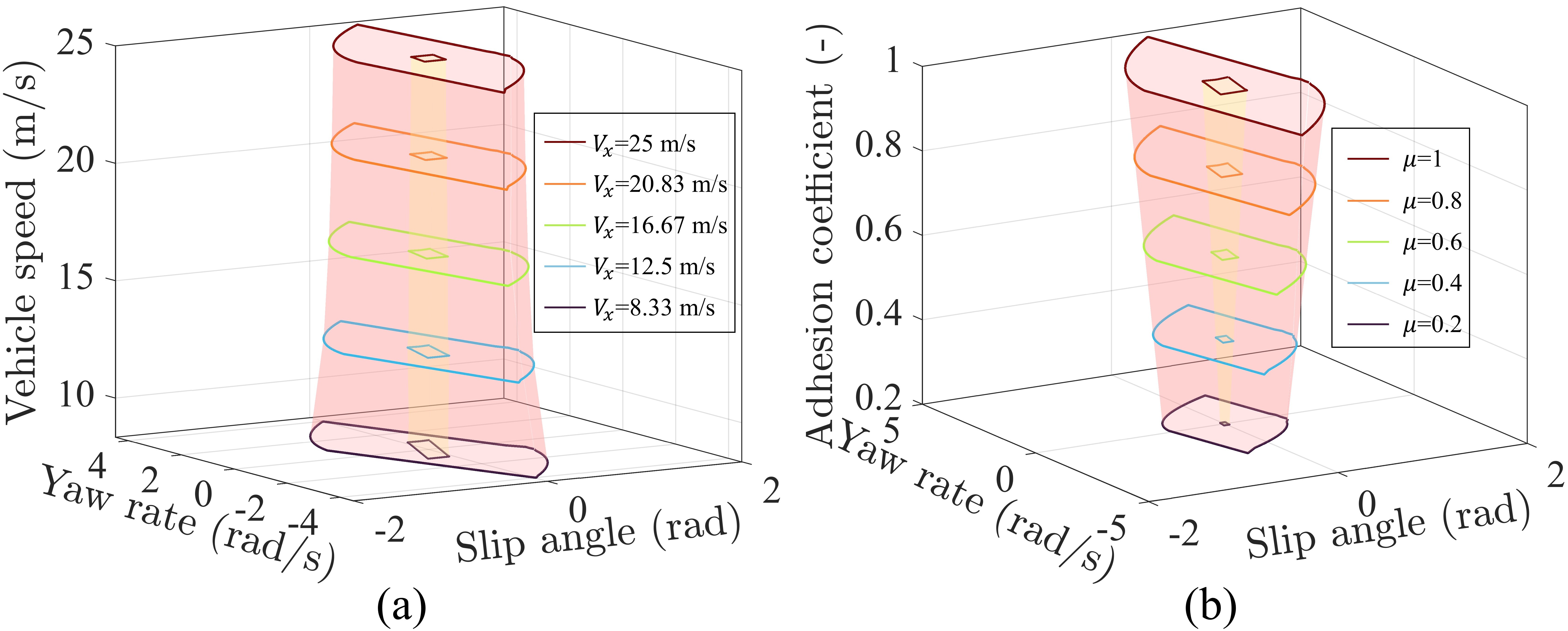}
  \caption{The drifting extended dual envelope with the change of different factors under the influence of control input is considered. (a) Vehicle speeds. (b) Road adhesion coefficients. }\vspace{-20pt}
  \label{The drifting extended dual envelope with the change of vehicle speeds under the influence of control input is considered.}
\end{figure}

Fig. \ref{The drifting extended dual envelope with the change of vehicle speeds under the influence of control input is considered.}(a) and Fig. \ref{The drifting extended dual envelope with the change of vehicle speeds under the influence of control input is considered.}(b) show the extended dual envelope surfaces for vehicle drifting under different vehicle speeds and road adhesion coefficients with bounded control inputs. The envelope size varies across operating conditions. As vehicle speed increases, the extended dual envelope gradually shrinks. In contrast, as the road adhesion coefficient increases, the extended dual envelope expands, which is consistent with saddle point trends in conventional vehicle dynamics.
\vspace{-10pt}
\section{Drifting Controller Design}\label{section:5}
% This section describes the design of the drifting controller structure. The proposed control strategy regulates the yaw rate and sideslip angle to track reference values by manipulating front wheel steering angles, rear longitudinal forces and additional yaw moments, thereby enabling accurate drift control.

% For a given vehicle speed and road adhesion condition, the vehicle sideslip angle and yaw rate may enter a region associated with instability and loss of control. In this situation, appropriate control inputs can drive the vehicle state toward the saddle point. The NMPC controller incorporates the extended dual envelope developed in the previous section as a constraint and includes penalty terms to reduce state errors and improve drift control accuracy.

%Based on the extended dual-envelope constraints derived in the previous section, an NMPC controller is designed to track the reference drift equilibrium while keeping the vehicle state within the controllable drifting region. The controller regulates the yaw rate and sideslip angle by optimizing the front wheel steering angle, rear longitudinal force, and additional yaw moment.

\subsection{NMPC Controller Design}
The NMPC controller is designed to compute an optimal solution for stable tracking and controllable drifting using a 3DOF vehicle model, while jointly optimizing vehicle states and actuator inputs over a finite prediction horizon. By combining (1), (2) and (3), an equivalent form of the vehicle dynamic equations can be obtained:
\begin{equation}
\dot{x}(t) = f(x(t),u(t)),
\end{equation}
where the state vector is $x = [e, \Delta \psi, V_x, \beta, r]$, and the control input is $u = [\delta, F_{xr}, \Delta M_z]$. The NMPC controller outputs the control command at the current sampling instant for real time closed loop execution. The optimization problem is implemented in CasADi, and the sequential quadratic programming is used as the solver. The initial reference drift equilibrium point is computed using through (1) and (3). At each sampling instant, the controller receives the current state and the applied input from the previous sampling instant, predicts future states, and solves for the optimal control sequence. 

To ensure that the current vehicle state and the predicted states satisfy the physical characteristics of the vehicle, Equation (19) is discretized using the fourth order Runge Kutta method. The resulting discrete time model is given below:
\begin{equation}
x_{k+1} = f(x_k,u_k,dt).
\end{equation}\vspace{-10pt}
\subsection{Multiple Constraints}
Constraints are key components of the NMPC controller because they ensure that vehicle states and control inputs remain within prescribed bounds, thereby supporting reliable controller operation. In Section \ref{section:4}, an extended dual envelope is constructed for vehicle states during drifting, which restricts predicted states to remain within the envelope. Additional constraints are imposed on the control inputs to ensure that the control actions remain within safe limits.\vspace{-10pt}
\subsubsection{Extended Dual Envelope Constraints}
The extended dual envelope constraints consists of an inner envelope and an outer envelope. The region inside the inner envelope is defined as a steady state control region, which is intended to prevent the NMPC solver from converging to an undesired local optimum that inhibits drifting. The outer envelope constrains drift motion, mitigates instability and loss of control, and supports stabilization of vehicle attitude.

For the outer envelope, the boundary is treated as the limit within which the vehicle state can be driven back under controller action. Therefore, a hard constraint is imposed on the outer envelope, meaning that predicted trajectories are not allowed to violate this boundary within the prediction horizon. Considering that the initial vehicle state lies in the stable region and should avoid entering the inner envelope region during drifting, the inner envelope is imposed as a soft constraint. The inequality forms of these constraints are given in (14) to (17). The above constraints are computed offline to generate a lookup table that provides bounds for the NMPC constraints.\vspace{-10pt}

\subsubsection{Control Inputs Constraints}
Constraints are also imposed on the control inputs. Vehicle actuators limit the allowable front wheel steering angle. The rear wheel longitudinal force is constrained to remain within the available friction limits. In addition, $\Delta M_z$ is constrained to satisfy actuator limits and drift safety requirements. The corresponding constraints are given by:
\begin{equation}
\left\{
\begin{aligned}
\delta_{\min} &\le \delta_{k} \le \delta_{\max};\\
-\mu F_{zr} &\le F_{xr k} \le \mu F_{zr};\\
\Delta M_{z\min} &\le \Delta M_{z k} \le \Delta M_{z\max}.
\end{aligned}
\right.
\end{equation}\vspace{-10pt}
\subsection{Objective Functions}
The NMPC controller improves drift control accuracy by tracking the reference targets while enforcing the above constraints. The objective function is defined as follows:
\begin{equation}
J(x_k,u_k)
=\sum_{k=1}^{N_p}\
\left\|x_k - x_{\mathrm{ref},k}\right\|_{Q}^{2}
+\sum_{k=0}^{N_c-1}\left\|u_k\right\|_{R}^{2},
\end{equation}
where $N_p$ and $N_c$ denote the prediction horizon and control horizon, respectively, with $N_p > N_c$. The matrices $Q$ and $R$ are the weight matrices in the NMPC cost function.
\vspace{-10pt}
\subsection{DDEV Torque Distribution}
The above NMPC controller uses rear wheel longitudinal force and additional yaw moment as control inputs. To form a closed loop system, these control inputs must be converted into front and rear wheel driving torques.

The rear wheel driving torque can be computed from $F_{xr}$, which is expressed as follows:
\begin{equation}
T_{rl}=T_{rr}=\frac{F_{xr}R_e}{2},
\end{equation}
where $R_e$ is the effective radius of the tire.

Because the rear tires remain saturated during drifting, allocating additional yaw moment through the rear wheel can be inefficient. Therefore, the additional yaw moment is applied through the front wheels. The corresponding front wheel torque is given as follows:
\begin{equation}
T_{fl}= \frac{\Delta M_z R_e}{d \cos\delta}, \quad T_{fr}= -\frac{\Delta M_z R_e}{d \cos\delta},
\end{equation}
% \begin{equation}
% \left\{
% \begin{aligned}
% T_{fl}&= \frac{\Delta M_z R_e}{d \cos\delta};\\
% T_{fr}&= -\frac{\Delta M_z R_e}{d \cos\delta},
% \end{aligned}
% \right.
% \end{equation}
where $d$ is the track width of the vehicle.\vspace{-10pt}

\section{Experimental Validation and Analysis}\label{section:6}
In this chapter, the precision and robustness of the vehicle under the extended dual envelope constraints are validated using a HiL experimental platform. Controller parameters are listed in Table \ref{tab:parameters}. Steady state drifting tests are conducted using two algorithms: (a) NMPC with the extended dual envelope constraints (Our Controller), and (b) NMPC without the envelope constraints (No Envelope Controller). To ensure a fair comparison, identical weights and horizons are used for both controllers.
\begin{table}[t]
\centering
\caption{Control Parameters}
\label{tab:parameters}
\begin{tabular}{c c c c}
\toprule
Symbol & Value & Symbol & Value \\
\midrule
 $N_p$  &  15 
 &  
 $Q$  & 
 $\mathrm{diag}(2400,\,4500,\,300,\,5000,\,1600)$  \\
 $N_c$  & 10 
 &  
 $R$  & 
 $\mathrm{diag}(1000,\,800,\,600)$  \\
$\delta_{\min}$ &  - 0.5 
&
$\Delta M_{z\min}$  & -3500\\
$\delta_{\max}$ &  0.5 
&
$\Delta M_{z\max}$  & 3500\\
\bottomrule
\end{tabular}
\end{table}
\vspace{-10pt}
\subsection{Drift Test Platform}

\begin{figure}[tbp]
  \centering
  \includegraphics[width=\linewidth]{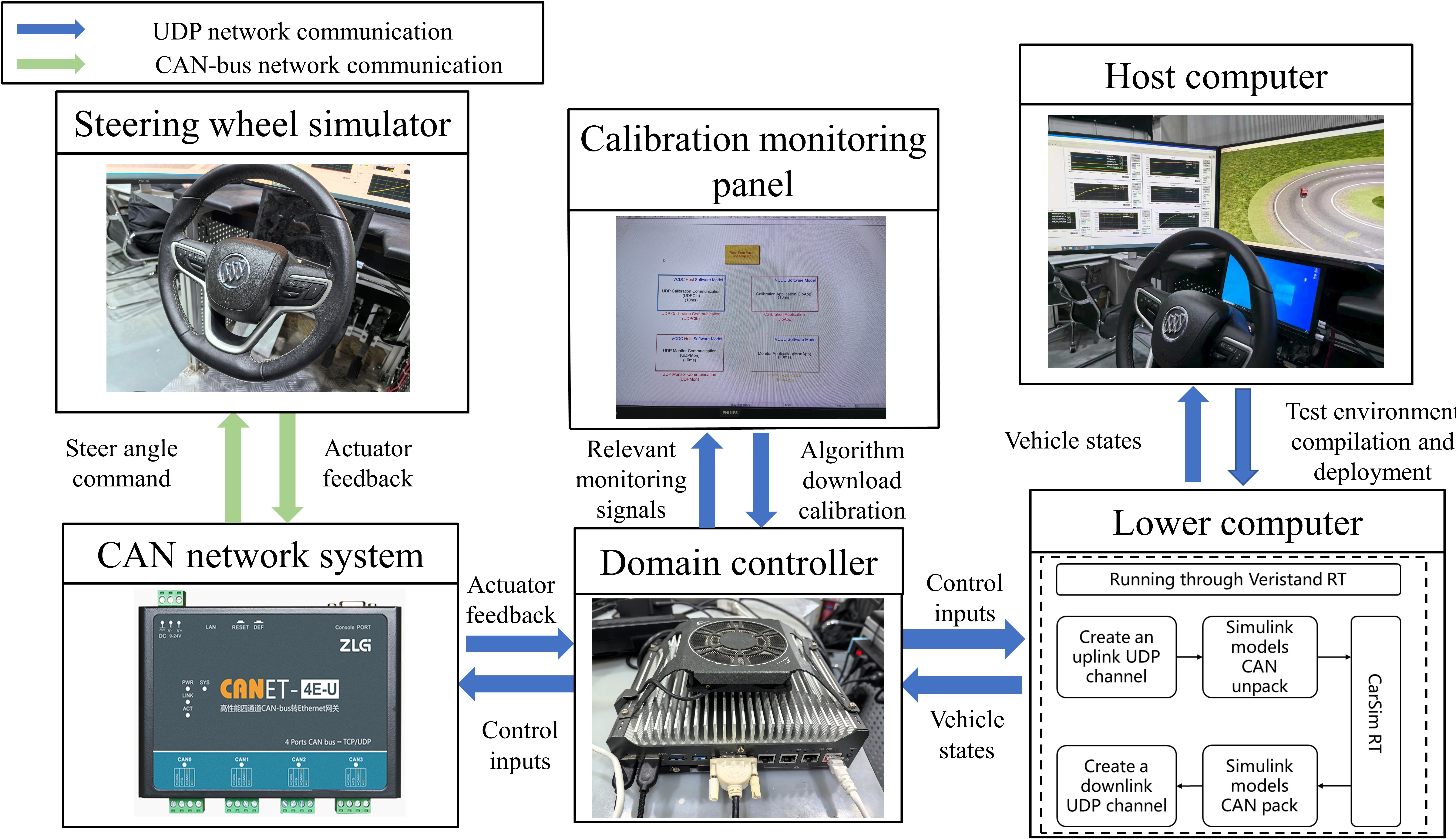}
  \caption{Hardware in the loop platform.}\vspace{-10pt}
  \label{Hardware in the loop platform.}
\end{figure}

As shown in Fig. \ref{Hardware in the loop platform.}, the domain controller executes the control algorithm. The domain controller uses an Intel Core i9 13900 processor to perform the computations required by the proposed method. The calibration monitoring panel is an integrated toolchain used to download control algorithms to the controller and perform parameter calibration and signal monitoring. The host computer is used to deploy the vehicle test environment. The real time version of CarSim 2019.1 is deployed on a target computer running VeriStand 2018 SP1 as DLL files, and the target computer transmits vehicle state data to the host computer. Communication among these components is implemented using UDP network. The steering wheel simulator and the CAN network subsystem communicate through the CAN. The steering wheel simulator emulates steering wheel input during drifting. To enable signal conversion between the CAN and UDP, a ZLG CANET-4E-U device is used. The CAN bus bit rate is 500 kb/s. The sampling time for all control algorithms is 20 ms. 
\vspace{-10pt}
\subsection{Steady-State Drift Test}

\begin{figure}[t]
  \centering
  \includegraphics[width=1\linewidth]{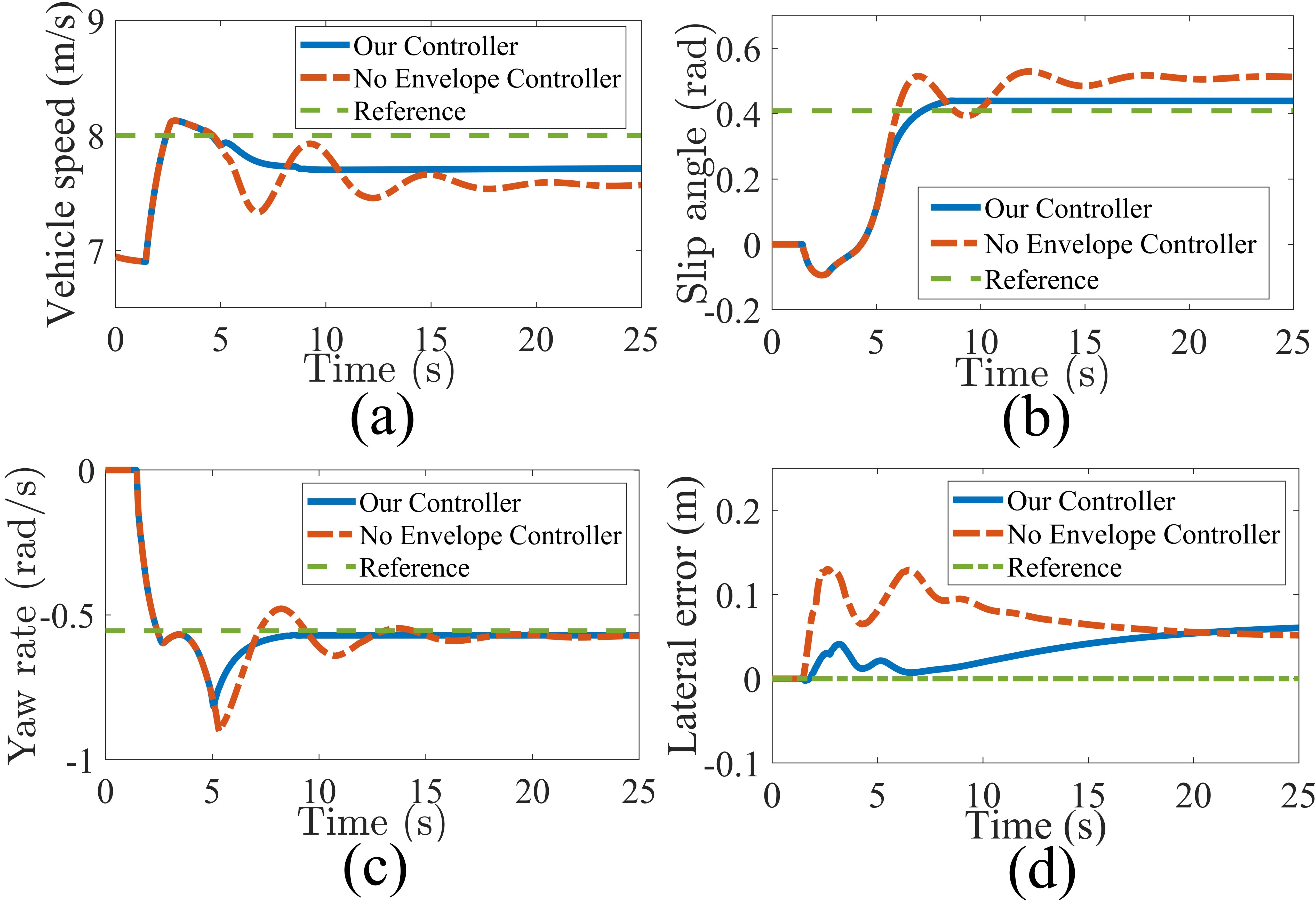}
  \caption{Comparison of state quantity changes in the drift process. (a) Vehicle speeds. (b) Slip angles. (c) Yaw rates. (d) Lateral errors.}\vspace{-15pt}
  \label{Comparison of state quantity changes in the drift process. (a) Vehicle speeds. (b) Slip angles. (c)Yaw rates. (d)Lateral errors.}
\end{figure}

\begin{figure}[t]
  \centering
  \includegraphics[width=0.7\linewidth]{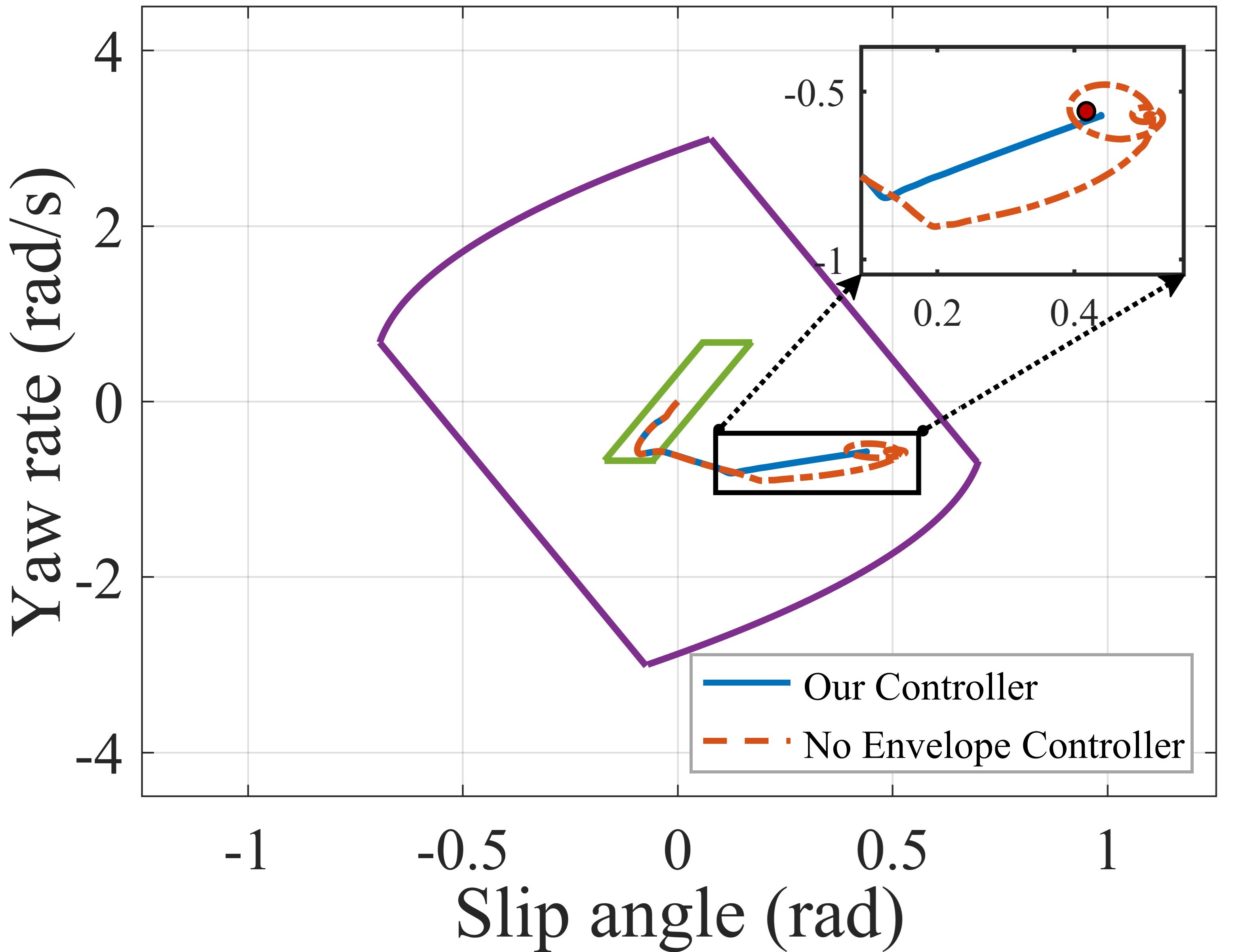}
  \caption{Constraint effects under controllers with and without the envelope.}\vspace{-15pt}
  \label{Constraint effects under controllers with and without the envelope.}
\end{figure}

\begin{figure}[t]
  \centering
  \includegraphics[width=1\linewidth]{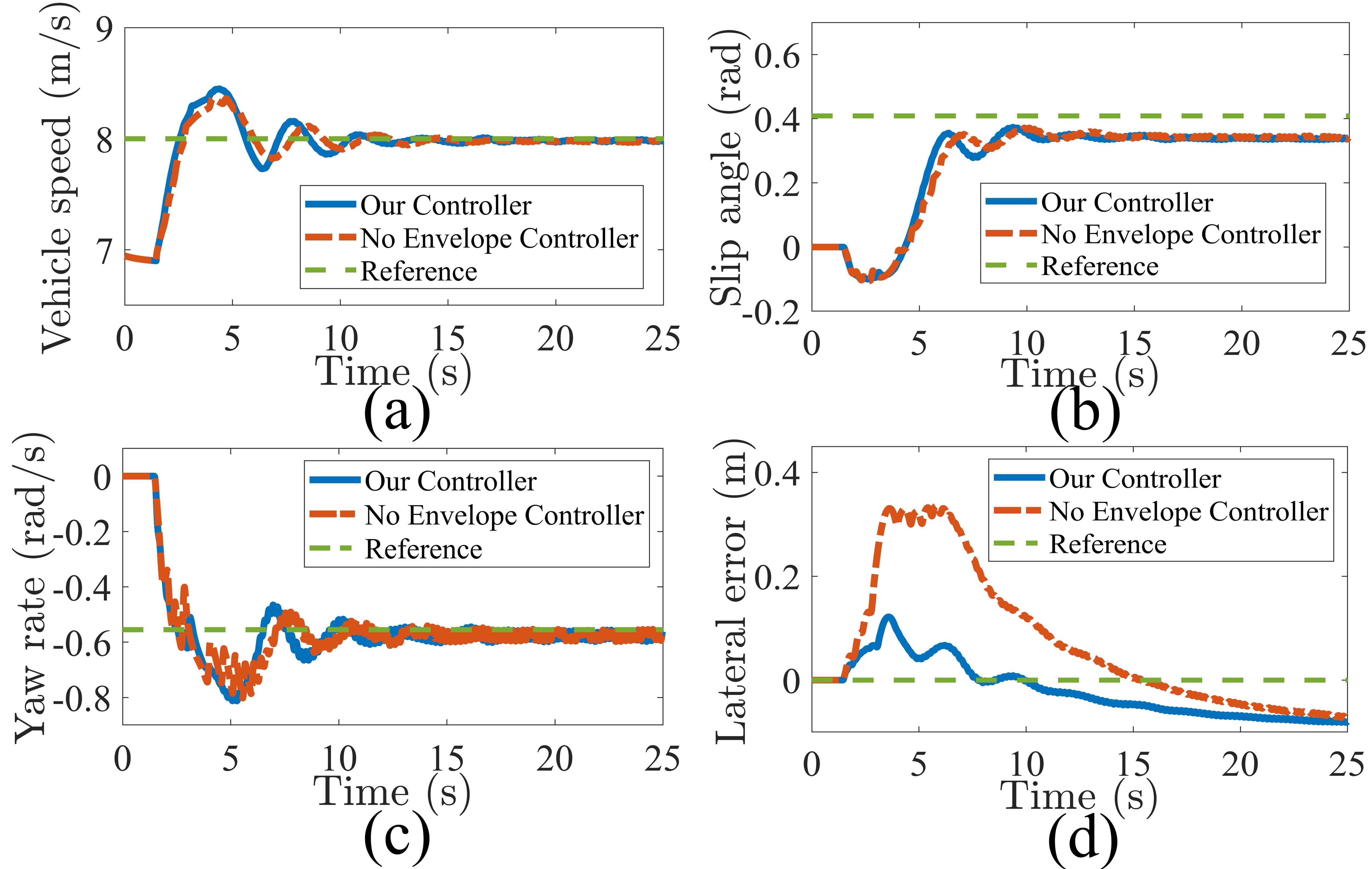}
  \caption{Comparison of state quantity changes in the drift process when $\mu=0.60$. (a) Vehicle speeds. (b) Slip angles. (c)Yaw rates. (d)Lateral errors.}\vspace{-15pt}
  \label{Comparison of state quantity changes in the drift process when.}
\end{figure}

\begin{figure}
  \centering
  \includegraphics[width=0.65\linewidth]{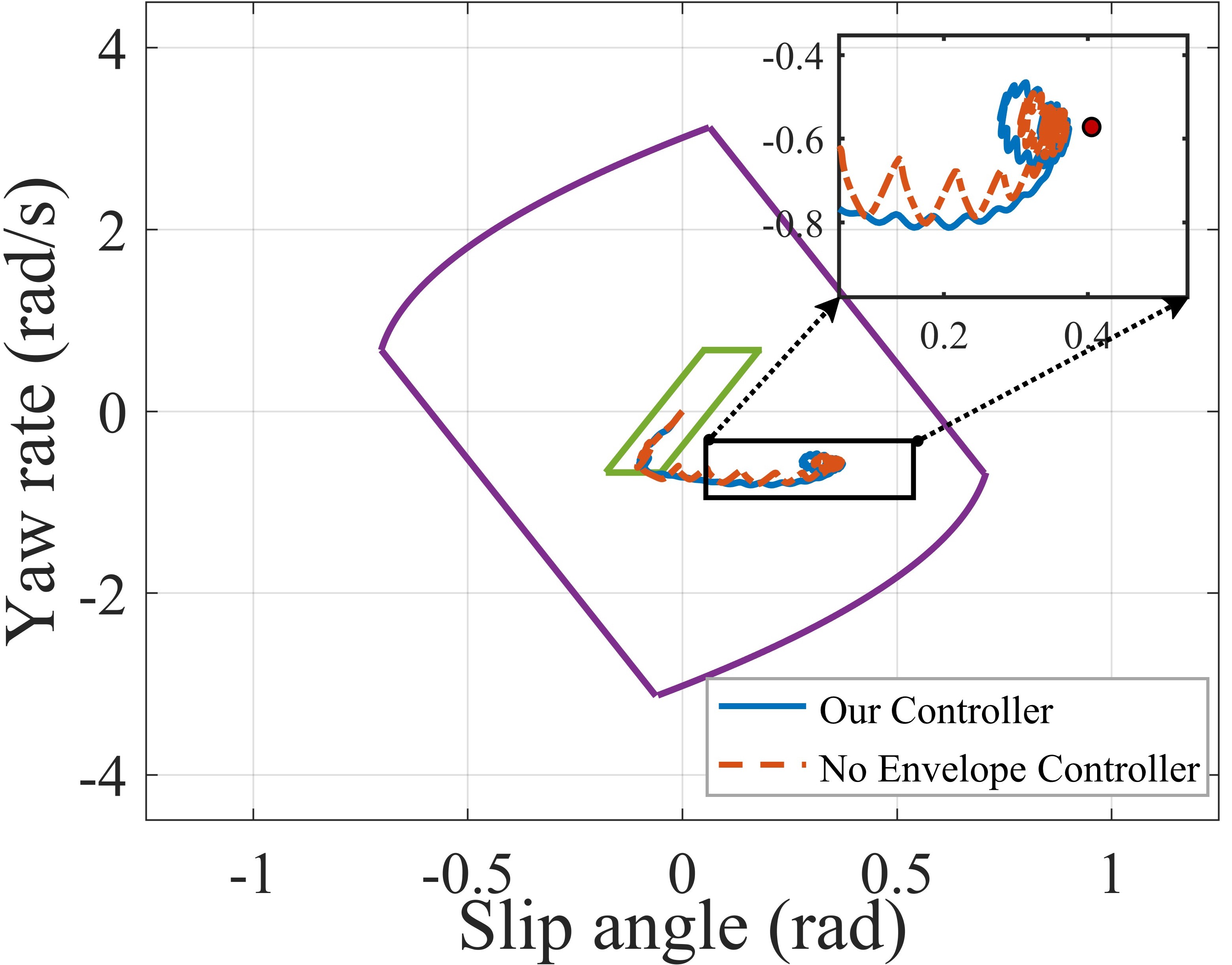}
  \caption{Constraint effects under controllers with and without the envelope when $\mu=0.60$.}\vspace{-20pt}
  \label{Constraint effects under controllers with and without the envelope when}
\end{figure}
This experiment evaluates vehicle state performance during steady state drifting with and without envelope constraints. The steady state drift test is conducted at a road adhesion coefficient of 0.55, a radius of 14.4 m and a vehicle speed of 28.8 km/h (8 m/s). The time histories of the vehicle state variables are shown in Fig. \ref{Comparison of state quantity changes in the drift process. (a) Vehicle speeds. (b) Slip angles. (c)Yaw rates. (d)Lateral errors.}. The term Reference denotes the desired steady state drift values of the corresponding state variables. As shown in Fig. \ref{Comparison of state quantity changes in the drift process. (a) Vehicle speeds. (b) Slip angles. (c)Yaw rates. (d)Lateral errors.}, vehicle speed, sideslip angle, and yaw rate track the reference more closely than in the case without the envelope constraints. Although the lateral error under the envelope constraints shows a gradual increasing trend relative to the case without the envelope constraints, the lateral error is smaller during the initial drifting phase. The final steady states errors are summarized in Table \ref{States Errors after Steady Drift Stabilization}. Compared with the controller without the envelope constraints, the proposed controller reduces the steady state errors of vehicle speed by 33.07\%, sideslip angle by 71.18 \%, and yaw rate by 31.27\%. The results indicate that the proposed controller with the extended dual-envelope constraints outperforms the controller without the envelope constraints. 

\begin{table}[t]
\centering
\caption{State Errors After Steady Drift Stabilization}
\label{States Errors after Steady Drift Stabilization}
\setlength{\tabcolsep}{2.8pt}
\renewcommand{\arraystretch}{0.95}
\footnotesize
\begin{tabular}{l c c c}
\toprule
Controller & Speed Error & Slip-Angle Error & Yaw-Rate Error \\
           & (m/s)       & (rad)            & (rad/s)        \\
\midrule
Our controller          & -0.2938 & 0.0302 & -0.0156 \\
No envelope controller  & -0.4390 & 0.1048 & -0.0227 \\
\bottomrule
\end{tabular}\vspace{-10pt}
\end{table}

A more detailed illustration of the constraint effects is provided in Fig. \ref{Constraint effects under controllers with and without the envelope.}. The red markers indicate saddle points. The blue solid curve shows the response of our controller, whereas the red dashed curve shows the response of no envelope controller. The proposed controller approaches the saddle point more smoothly. In contrast, the controller without the envelope constraints exhibits larger oscillations and overshoot near the saddle point, and its final state remains farther from the saddle point, indicating lower accuracy.
\vspace{-15pt}
\subsection{Steady-State Drift Robustness Test}

This experiment evaluates controller robustness by comparing the extended dual envelope constrained controller with a controller that does not use the envelope constraints. Controller parameters are identical to those in the previous experiment, but the actual road friction coefficient is 0.60, which differs by 0.05 from the value assumed in the constraint design. As shown in Fig. \ref{Comparison of state quantity changes in the drift process when.}, both controllers achieve steady state drifting, although small oscillations are observed, attributed to the selected state weights. Fig. \ref{Comparison of state quantity changes in the drift process when.}d shows that the proposed controller yields a smaller tracking error than the controller without the envelope constraints. The maximum error of the proposed controller is 0.1219 m, whereas the maximum error of the controller without the envelope constraints is 0.3354 m, corresponding to a peak error reduction of approximately 63.66 \%. The envelope constrained controller achieves higher tracking accuracy than the controller without the envelope constraint under friction coefficient mismatch, thereby demonstrating improved robustness during drifting.

Fig. \ref{Constraint effects under controllers with and without the envelope when} compares the responses with and without the envelope constraints under road adhesion coefficient deviation. Although the final steady state drift performance is similar in both cases, the yaw rate oscillations are more effectively suppressed by the proposed controller during tracking to the reference. This observation supports the robustness of the proposed controller under envelope constraints during drifting.\vspace{-10pt}

\section{CONCLUSION}\label{section:7}
% This paper derives a multifactor saddle point location expression that incorporates control inputs by analyzing the effects of front wheel steering angle and additional yaw moment on saddle point dynamics. Using this expression, an extended dual envelope suitable for vehicle drifting is developed by accounting for front and rear tire saturation and the convergence tendency of each state point toward the saddle point under control action. The resulting extended dual envelope is incorporated into a state constrained NMPC drift controller. Experimental results show that, compared with the controller without the envelope constraints, the proposed controller effectively suppresses oscillations during drifting, improves control accuracy, and provides increased robustness. Future work will investigate more precise control strategies.

This paper develops an extended dual-envelope-constrained NMPC method for DDEVs drifting. A multifactor saddle-point location expression is first derived by considering the effects of front wheel steering angle and additional yaw moment on saddle-point dynamics. Based on this, an extended dual envelope is constructed to characterize the controllable drifting region and is incorporated into the NMPC controller as a state constraint. HiL experimental results show that the proposed method improves drift stability, tracking accuracy, and robustness compared with the controller without envelope constraints. In steady-state drifting, the errors of vehicle speed, sideslip angle, and yaw rate are reduced by 33.07\%, 71.18\%, and 31.27\%, respectively, while under road-friction mismatch the peak tracking error is reduced by 63.66\%. Future work will investigate adaptive envelope design and higher-performance drift control strategies.

\bibliographystyle{IEEEtran}
\normalem
\bibliography{IEEEabrv,references}

\end{document}